 \documentclass[final,5p,times,twocolumn]{elsarticle}

\usepackage{amssymb}
\usepackage{subfigure}
\usepackage{amsmath}
\usepackage[linesnumbered, ruled]{algorithm2e}
\usepackage{url}
\usepackage{hyperref}

\journal{Nuclear Physics B}

\begin{document}

\begin{frontmatter}



\title{DRL-M4MR: An Intelligent Multicast Routing Approach Based on DQN Deep Reinforcement Learning in SDN}

\author[guet2]{Chenwei Zhao} 
\ead{zhaochenwei96@foxmail.com}

\author[guet2,guet2k]{Miao Ye\corref{cor1}} 
\ead{yemiao@guet.edu.cn}

\author[fujian]{Xingsi Xue} 

\author[pengcheng]{Jianhui Lv} 

\author[guet2,guet2k]{Qiuxiang Jiang} 

\author[guet3]{Yong Wang} 

\cortext[cor1]{Corresponding author}

\address[guet2]{School of Information and Communication, Guilin University of Electronic Technology, Guilin 541004, China}
\address[guet2k]{Guangxi Key Laboratory of  Wireless Wideband Communication and Signal Processing, Guilin University of Electronic Technology, Guilin 541004, China}
\address[fujian]{Fujian Provincial Key Laboratory of Big Data Mining and Applications, Fujian University of Technology, Fuzhou, Fujian, 350118, China}
\address[pengcheng]{Peng Cheng Lab. Shenzhen, Guangdong, 518038, China}
\address[guet3]{School of Computer Science and Information Security, Guilin University of Electronic Technology, Guilin 541004, China}

\begin{abstract}
Traditional multicast routing methods have some problems in constructing a multicast tree, such as limited access to network state information, poor adaptability to dynamic and complex changes in the network, and inflexible data forwarding. To address these defects, the optimal multicast routing problem in software-defined networking (SDN) is tailored as a multi-objective optimization problem, and an intelligent multicast routing algorithm DRL-M4MR based on the deep Q network (DQN) deep reinforcement learning (DRL) method is designed to construct a multicast tree in SDN. First, the multicast tree state matrix, link bandwidth matrix, link delay matrix, and link packet loss rate matrix are designed as the state space of the DRL agent by combining the global view and control of the SDN. Second, the action space of the agent is all the links in the network, and the action selection strategy is designed to add the links to the current multicast tree under four cases. Third, single-step and final reward function forms are designed to guide the intelligence to make decisions to construct the optimal multicast tree. The experimental results show that, compared with existing algorithms, the multicast tree construct by DRL-M4MR can obtain better bandwidth, delay, and packet loss rate performance after training, and it can make more intelligent multicast routing decisions in a dynamic network environment.
\end{abstract}

%


\end{frontmatter}


\section{Introduction}\label{sec:intro}
In many applications, such as video websites, teleconferencing, and multi-copies of cloud storage systems, if a unicast routing strategy is adopted to transmit the same data to multiple destinations, the same transmission link of the redundant data consumes extra network resources, making it easier for the transmission link to produce problems such as network congestion and increased traffic consumption. Therefore, in the face of these "one-to-many" scenarios, the multicast technique can better solve these problems. In multicast, when the same data packet is sent, the source node sends only one of the data to multiple destination nodes, and the destination node receives a copy of the sent data \cite{RN1}, thus redundancy consumption can be reduced during data transmission \cite{RN2}. It is necessary to construct a Steiner tree (ST) or a multicast tree from the source node to the destination nodes to meet the quality of service (QoS) requirements to determine the multicast routing path. The Steiner tree problem is an NP-complete problem \cite{RN3}. The design and implementation of efficient and reasonable multicast routing are inseparable from the timely and convenient acquisition of various network status information, including remaining link bandwidth, link packet loss rate, and link delay. However, traditional methods of obtaining network state information, such as open shortest path first (OSPF) and internal gateway routing protocol (IGRP) have complicated configurations and steps for measuring the network state. With an increasing node scale, the cost of obtaining network state information will increase rapidly, and the network convergence will be slow \cite{RN4}. Compared with the traditional network measurement method, software-defined networking (SDN) is a new network architecture that abstracts the network system model and provides a more flexible and convenient way to measure and obtain network link information (NLI). The SDN decouples the control plane from the data plane and leverages centralized network control to make it easy to configure and manage the network by programming on the controller \cite{RN5}. In the SDN architecture, the controller communicates with the data plane through a unified and open southbound interface (SBI) for operations such as routing flow table delivery and obtaining network status information \cite{RN6}. Routing is a fundamental function of the network. Inefficient routing decisions will result in increased data transfer delays and link loads and will affect the overall performance of the SDN. The centralized control and programming of the SDN controller enable multicast routing algorithms to efficiently utilize and reprogram network resources \cite{RN7}. Therefore, optimizing routing in SDN is an important study \cite{RN8}. There has been much research on SDN unicast routing, but little research on optimizing SDN multicast routing performance.

Currently, some approaches to SDN multicast routing apply only the classic multicast tree methods. In Paper \cite{RN9} the shortest path tree (SPT) algorithm is used to construct multicast trees, but the SPT algorithm cannot make full use of the network resources. Additionally, many heuristic algorithms for constructing multicast trees have been proposed, such as the KMB (Kou, Markousky, and Berman proposed) algorithm \cite{RN10} and the minimum cost path heuristic (MPH) algorithm \cite{RN11}, which can solve the problem in polynomial time but cannot obtain the optimal ST. Therefore, there are many intelligent optimization methods, such as the ant colony optimization (ACO) algorithm \cite{RN12}, genetic algorithm (GA) \cite{RN13}, particle swarm optimization (PSO) \cite{RN14}, etc. can solve the ST problem well. But when multiple multicast trees need to be constructed at the same time, greater computing capacity is needed, which brings a computational burden to the controller and leads to a longer solving time.

In recent years, many studies have applied reinforcement learning (RL) methods to routing problems. RL is a goal-oriented learning method that automatically processes problems through sequential decision-making. RL does not require complete modeling of the environment, it emphasizes agent learning through direct interaction with the environment. The environment feeds back the reward and the agent seeks to optimize the goal to maximize the reward. Compared with the heuristic algorithm, RL can provide routing policies quickly after training, instead of requiring online recalculation each time the multicast tree is constructed. The RL algorithm achieves significant performance improvements in routing (e.g., delay, loss, throughput, and other metrics) and can adapt more intelligently to dynamic environments than traditional static routing algorithms. \cite{RN15,RN16}. The solution in References \cite{RN17,RN18} still need to rely on the traditional shortest path algorithm and the agent needs to be retrained when the network environment changes. References \cite{RN19,RN20} use only hop to construct the multicast tree, without considering other network state information.

Compared with the existing work in this field, we propose an intelligent multicast routing algorithm, which utilizes deep reinforcement learning methods to construct multicast trees in SDN by using four matrices as state spaces (DRL-M4MR). The algorithm abstracts the construction of a multicast forwarding path as a Markov decision process (MDP). The state space, action space, and reward function are designed and multiple state information of network links are used to construct multicast routes. The DRL-M4MR takes the available link bandwidth, link delay, link packet loss rate matrix, and multicast tree state matrix as the state space of the agent. The agent uses all links in the network as action spaces, and the action selection policy is designed to add a link to the current multicast tree in four cases. The DRL-M4MR agent interacts with the environment to obtain rewards and punishments in three cases and learns from experience to calculate an optimal multicast tree from the set of known source nodes to destination nodes. After training, the agent can find the optimal multicast tree in the environment of network state change. Finally, the controller reversely traverses the multicast tree from the leaf nodes and installs the multicast routing flow table to the switch on the data plane through the southbound interface.

The contributions of this paper are as follows: 
\begin{enumerate}[1)]
\item Combining the programmable feature of SDN and the deep reinforcement learning method, three kinds of link-state information (available bandwidth, delay, and packet loss rate) are measured and used as factors to design the state space and reward function of the agent.
\item The matrix form of agent state space is designed, including multicast tree state matrix, remaining bandwidth matrix, delay matrix and packet loss rate matrix. In the state space design, all links in the network are used as outputs of fixed dimensions of the neural network, and actions correspond to links one by one, and an action strategy corresponding to the four cases is selected for each action. The single-step reward function and the final reward function are designed to guide agents to explore the multicast tree based on the information measured in the SDN network and to make better decisions to build the multicast tree.
\item The controller ignores redundant nodes by reversely traversing the multicast tree constructed by the agent from the destination node to the source node when installing the flow table. The flow entries on fork nodes are configured with multiple output actions to forward packets to different ports.
\item To improve the convergence efficiency and stability of DRL, double dueling DQN and prioritized experience replay (PER) are adopted. Additionally, decay $\epsilon$-greedy is adopted to better balance exploration and exploitation.
\end{enumerate}

The remainder of this paper is organized as follows. Related works are discussed in Section \ref{sec:relate}. The SDN multicast problem is described in Section \ref{sec:description}. The architecture and modeling of intelligent SDN multicast routing optimization are introduced in Section \ref{sec:architecture}. The design of the DRL-M4MR algorithm agent and the details of the algorithm are presented in Section \ref{sec:DRL-M4MR} to implement multicast routing. The evaluation and comparison results are provided in Section \ref{sec:experiments}. The conclusions and suggestions for future work are presented in Section \ref{sec:Conclusion}.

\section{Related works}\label{sec:relate}
In this section, we will discuss previous related studies that explored different techniques to improve multicast routing.

\textit{Classical algorithm}: The classical heuristic algorithms include the KMB algorithm, MPH algorithm, and the average distance heuristic (ADH) algorithm. The KMB algorithm is based on a minimum spanning tree and shortest path algorithm \cite{RN10}. First, the complete distance graph $G^\prime$ of graph $G$ is constructed. Then, the subgraph $G_1$ containing the destination node in $G^\prime$ and the minimum spanning tree $T_1$ of $G_1$ is found. $T_1$ is expanded to the subgraph $G^{\prime\prime}$ of G according to the shortest path in $G$. Finally, the minimum spanning tree of $G^{\prime\prime}$ is found, and the redundant codon nodes are deleted; that is, the final quasi-Steiner tree is obtained. The time complexity of the KMB algorithm is $O\left(mn^2\right)$, and the approximate optimal Steiner tree is generally obtained. The MPH algorithm adds nodes from the shortest path to the spanning tree each time until all destination nodes are included. This algorithm has a similar problem to the KMB algorithm, and the complexity is $O\left(mn^2\right)$. The ADH \cite{RN21} algorithm constructs a Steiner tree by connecting two subtrees with the shortest path. The time complexity of the ADH algorithm is $O\left(n^3\right)$.

Many subsequent algorithms are based on classical heuristic algorithm improvement. The delay-constrained minimum cost path heuristic (DCMPH) algorithm \cite{RN22} considers the delay constraint condition when generating the lowest cost Steiner tree according to the MPH algorithm. If the delay threshold is met, the multicast path is directly added to the multicast tree. If the delay threshold is not met, the multicast path is added to the multicast tree through the minimum delay path generated by the minimum spanning tree. However, the algorithm requires additional operations to eliminate the introduced loops. The key node-based minimum cost path heuristic (KBMPH) algorithm \cite{RN23} by finding the key node with more shortest path, if the path to be added to the multicast tree passes through more than one key node then correct the cost of the path, and then do a comparison whether to join the tree. This algorithm will increase the number of shared nodes, but the complexity is $O\left(n^3\right)$. These algorithms all have high computational complexity, because they rely on classical algorithms, and also have problems of slow convergence and difficulty in obtaining optimal solutions when the network scale is large.

Kotachi et al. \cite{RN24} regard determining multicast tree routing as an integer linear programming (ILP) problem to minimize the total number of flow entries. Latif et al. \cite{RN25} utilized ILP to optimize multicast trees and improve fabric utilization in CLOS networks. Touihri et al. \cite{RN26} expressed the computation of multicast trees as a mixed integer linear programming (MILP) to maximize the minimum remaining bandwidth and the number of links in the SDN-based CamCube Data Center Network. Linear programming requires considerable computation, and its constraints are often designed for specific scenarios.

\textit{Intelligent optimization algorithm}: Hassan et al. \cite{RN12} used the ACO algorithm to solve the problem of multicast routing based on total cost, delay, and hop count, which required considerable time to solve. Zhang et al. \cite{RN13} introduced a new crossover mechanism, the leaf crossover genetic algorithm (LCGA), to obtain better QoS multicast routing. Although LC has brought performance improvement, it takes a long time to implement. Shakya et al. \cite{RN14} adopted the bi-velocity particle swarm optimization (BVDPSO) algorithm with a two-speed strategy to solve the multicast routing problem so that the algorithm retained the global search ability of the original particle swarm and the fast convergence speed. Sahoo et al. \cite{RN27} embedded the PSO algorithm into the bacterial foraging (BFA) algorithm to construct QoS multicast routing according to end-to-end delay, bandwidth, delay jitter, and cost; however, iteration time is expensive. Murugeswari et al. \cite{RN28} proposed the multicast algorithm by combining the PSO algorithm and GA, which can obtain the tree of the minimum cost from the source node to the destination node according to the delay and bandwidth. These algorithms all face the same problem. When multiple multicast trees need to be built at the same time, more computational power is required. Therefore it is not suitable to compute a routing policy for each new flow in the controller \cite{RN8}.

\textit{Reinforcement learning method}: There has been some research on the reinforcement learning routing problem, and the general framework is as follows \cite{RN15}. The data plane is responsible for packet forwarding. On the control plane, the SDN controller is responsible for sensing the network topology, collecting network state information and installing flow tables. The management plane stores statistics and processes them into a form that can be used for RL. The knowledge plane uses this data to train the agent to find the best routing strategy \cite{RN29,RN30}. The controller then installs flow tables to the switch to complete routing tasks.

According to the differently designed action space, the routing algorithm of RL is divided into three types. (1) The agent selects a node from the neighboring nodes of the current routing node as the next switch for forwarding packets until the data is forwarded to the destination node \cite{RN16,RN31,RN32}. (2) The action space is the value of weights for each link in the network topology, and each action assigns different values to these weights. Finally, the shortest path is calculated based on the weights using the Dijkstra algorithm or Floyd algorithm \cite{RN18,RN33}. (3) The third type of action space is based on the k shortest paths calculated by the shortest path algorithm, and one path is chosen for each action \cite{RN17}. However, these methods predominantly consider the discussion of SDN unicast routing and cannot be directly used to solve the problem of multicast routing. 

Forster et al. \cite{RN34} used Q-learning in a wireless sensor network (WSN) to solve the multicast problem of the sink node. The information of neighbor nodes of the destination node and the current routing node was used as the state space. The action space decomposes actions into sub-actions to represent the selection of a node or a different neighbor node. After the reward calculates the reward for the entire action, backpropagation calculates the reward for each sub-action. Heo et al. \cite{RN19} use DQN to solve the multicast routing problem in SDN. A matrix with a special identifier is used to represent the current state of the network. However, only the hop count of the multicast tree is considered without taking into account the bandwidth, latency, packet loss rate, etc. in a dynamic network environment. Chae et al. \cite{RN20} proposed Advantage Actor-critic (A2C) to solve the multicast tree problem with dynamic network topology changes. The network topology, multicast tree and multicast set information as the state. The action is to select an edge that is not in the current tree and add it every time. If the tree reaches a destination node, a return value will be obtained. Only the length of the tree is used as a metric, other network information is not considered.

To make multicast routing better adapt to the dynamic network environment and make full use of network resources. In this paper, bandwidth, delay and packet loss rate are considered as the metrics, and the DRL method is adopted to construct the multicast tree in SDN.

\section{Description of SDN multicast problems}\label{sec:description}
\subsection{Problem formulation}\label{subsec:problem formulation}
\begin{figure}[ht]
	\centering
	\includegraphics[width=0.3\textwidth]{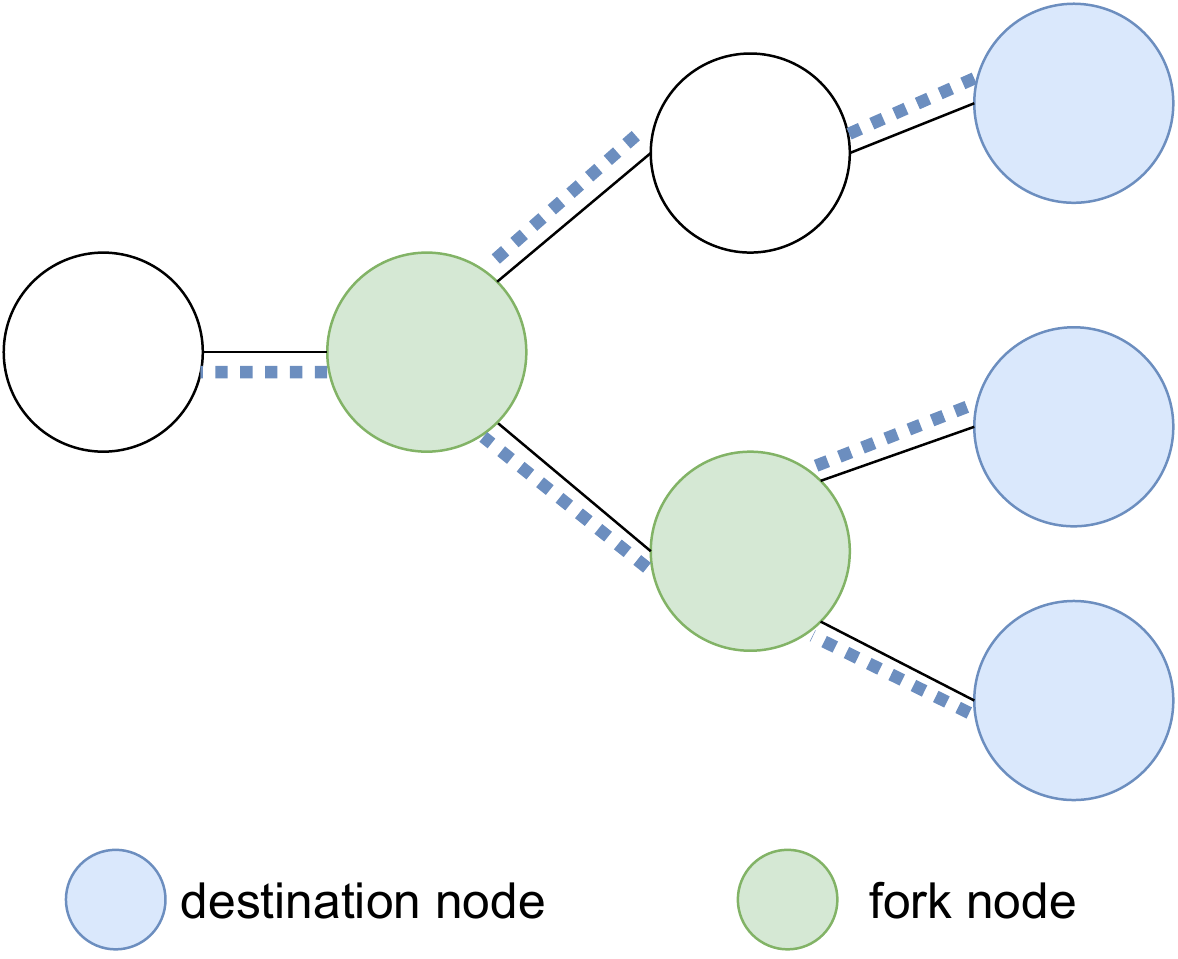}
	\caption{An example of a multicast tree.}
	\label{fig:multicast_tree}
\end{figure}

Multicast means that data packets are sent from a source node and received by a group of target nodes. When the data packets start from the source node, there is only one piece of data. Subsequently, a node (which can be called a fork node) replicates the data to nodes in multiple branches and then forwards it. The transmission path of the packet is a tree with the source node as the root and the destination node as the leaf node, as shown in Figure.\ref{fig:multicast_tree}. The topological tree structure is called a multicast tree (or Steiner tree). The optimization of a multicast tree involves finding a certain multicast tree to achieve optimal performance, usually called the minimum Steiner tree problem \cite{RN23}.

In a given network $G=\left(V,E\right)$, where $V$ is the nodes in graph $G$, $n=\left|V\right|$ denotes the number of nodes, $E$ is a collection of links in $G$, $e_{ij}\in E$ denotes the link between node $i$ and node $j$, $m=\left|E\right|$ denotes the number of links, and $S\subseteq V$ denotes the set of multicast nodes. The source node is denoted by $s\in S$, and $D=(d_1,d_2,\ldots)\in S$ denotes the set of destination nodes. The minimum Steiner tree of the source node is defined as finding the minimum span tree $T$ from $s$ to $D$ of $G$, which makes $S\in T$, $S\subseteq T \subseteq V$, and the cost of the tree minimum. Considering the characteristics of SDN, multicast trees are mainly measured in this paper from the following parameters:

$bw_{tree}$ is the average bandwidth of the multicast tree, which is the average of the minimum remaining bandwidth from the source node $s$ to each destination node $d$. It is defined according to Formula \ref{eq:bw_tree}.
\begin{equation}
	{f_1}(T) = bw_{tree} = average(\sum\limits_{d \in D} {\mathop {\min }\limits_{{e_{ij}} \in {p_{sd}}} (b{w_{ij}})} )
	\label{eq:bw_tree}
\end{equation}
where $bw_{ij}$ is the bandwidth from node $i$ to node $j$, and $p_{sd}\in T$ denotes the path from the source node $s$ to the destination node $d$ in the multicast tree $T$.

${delay}_{tree}$ is the cumulative delay of the multicast tree and represents the sum of the delays of all links in $T$, which is defined by the following Formula \ref{eq:delay_tree}.
\begin{equation}
	{f_2}(T) = delay_{tree} = \sum\limits_{{e_{ij}} \in T} {delay_{ij}}
	\label{eq:delay_tree}
\end{equation}
where ${delay}_{ij}$ denotes the delay from $i$ to $j$.

${loss}_{tree}$ is the packet loss rate of the multicast tree, which is calculated by Formula \ref{eq:loss_tree}.
\begin{equation}
	{f_3}(T) = loss_{tree} = 1 - \prod\limits_{{e_{ij}} \in T} {(1 - loss_{ij})}
	\label{eq:loss_tree}
\end{equation}
where ${loss}_{ij}$ denotes the packet loss rate of the link from $i$ to $j$.

The objective is to maximize $bw_{tree}$ and minimize ${delay}_{tree}$ and ${loss}_{tree}$. These objectives may be independent of each other. Therefore, the multicast problem under the SDN architecture considered to be solved in this paper is essentially a mathematical multi-objective optimization problem, as shown in Formula \ref{eq:F} below.
\begin{equation}
	\max F(T) = ({f_1}(T),{f_2}(T),{f_3}(T))
	\label{eq:F}
\end{equation}
where $T$ is a Steiner tree in $G$; and $F$ is the 3-dimensional target vector, which is the mapping of the decision variable $T$ to the 3-dimensional target space.

\section{SDN intelligent multicast routing architecture}\label{sec:architecture}
In this section, we introduce the overall structure of the intelligent SDN multicast routing method model. As illustrated in Figure \ref{fig:architecture}, the design of the intelligent multicast routing architecture includes the data plane, control plane, and management plane. Each component module and its function design are as follows:
\begin{figure}[ht]
	\centering
	\includegraphics[width=0.4\textwidth]{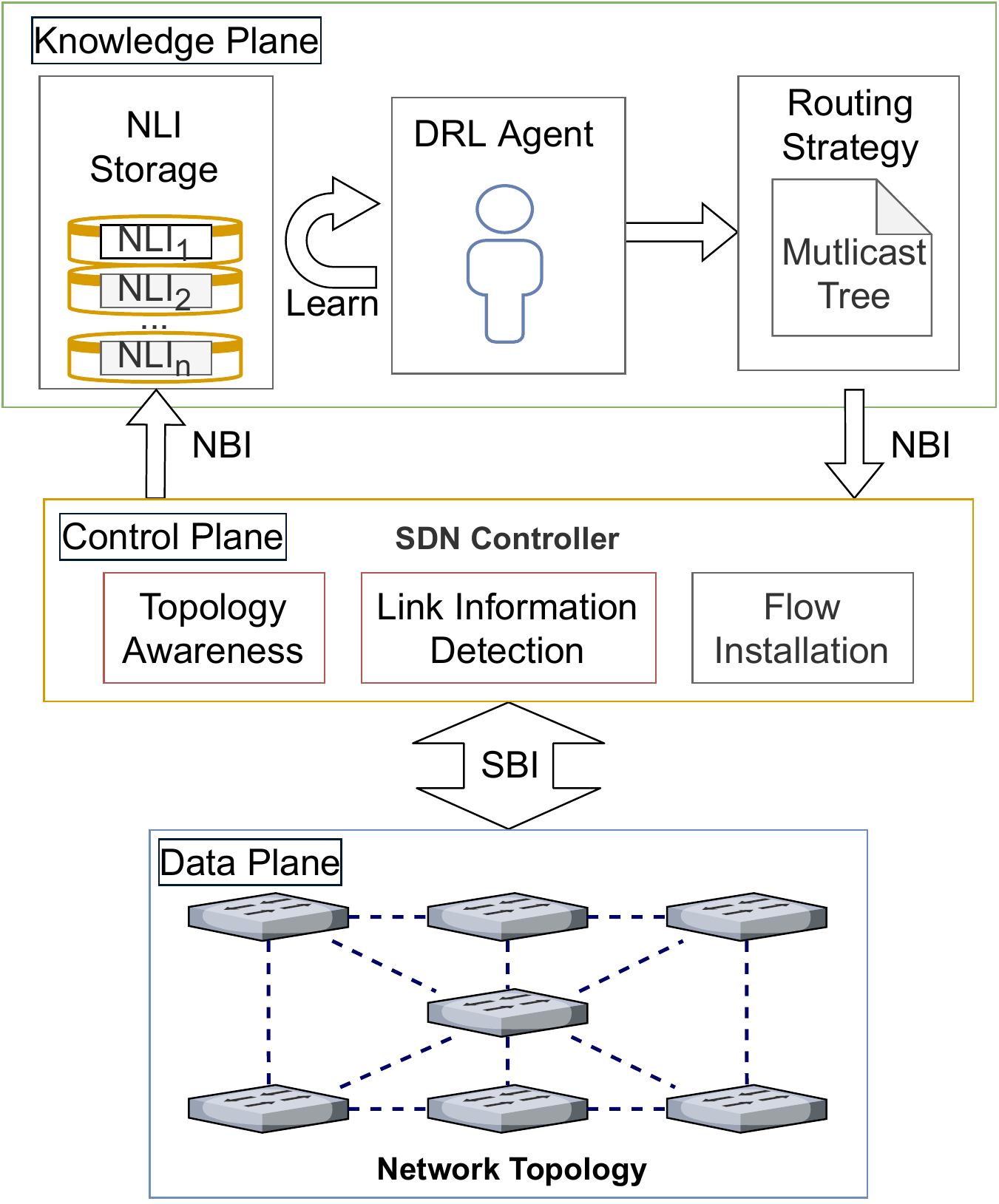}
	\caption{SDN intelligent multicast routing architecture.}
	\label{fig:architecture}
\end{figure}

\subsection{Data plane}\label{subsec:data_plane}
This plane consists of OpenFlow switches. The forwarding of data packets is achieved in this plane. Each SDN switch completes packet parsing and forwards packets from the input port to the output port based on matches in the internal flow table. In addition, the data plane can respond to requests from the control plane through the southbound interface. When forwarding multicast data, the switch matches the multicast address and forwards the data to one or more ports based on the multicast routing flow entries.

\subsection{Control plane}\label{subsec:control_plane}
This plane contains the centralized SDN controller. The controller is connected to each switch on the data plane and communicates via the SBI. In this way, the controller can control the switches centrally and easily gather information about the switches and links from a global view.

The controller provides three modules: \textit{Topology Awareness}, \textit{Link Information Detection}, and \textit{Flow Installation}. First, the \textit{Topology Awareness} module actively monitors the joining and leaving of switches, the addition, removal and modification of ports, and the addition and removal of links on the data plane. When these events occur, the controller sends link layer discovery protocol (LLDP) packets to the switches. The packets are propagated between topologies and back to the controller. The controller parses the LLDP data to obtain switch IDs, input and output port numbers and establishes the network topology based on the information.

Then, the link information is detected on the established network topology. The detected link information includes the remaining link bandwidth $bw_{ij}$, the link delay $delay_{ij}$, and the link packet loss rate $loss_{ij}$. The controller sends a request for port information to the data plane every t seconds. And the reply is parsed to obtain statistics on the number of sent bytes $t_b$, the received bytes $r_b$, the sent packets $t_p$, the received packets $r_p$, and the validity time $t_{dur}$ for each port $t_b$. The two statistics calculate the instantaneous throughput $bw_u$, as shown in Formula \ref{eq:bw_u}. The remaining bandwidth $bw_{ij}$ is the difference between the maximum bandwidth $bw_c$ and $bw_u$ of the current link, which employs Formula \ref{eq:bw_ij}.

\begin{equation}
	bw_u = \frac{{\left| {\left( {{t_{b2}} + {r_{b2}}} \right) - \left( {{t_{b1}} + {r_{b1}}} \right)} \right|}}{{{t_{dur2}} - {t_{dur1}}}}
	\label{eq:bw_u}
\end{equation}
\begin{equation}
	bw_{ij} = bw_c - bw_u
	\label{eq:bw_ij}
\end{equation}

The link packet loss rate $loss_{ij}$ calculates the maximum value from $i$ to $j$ and from $j$ to $i$. The formula is shown as follows:
\begin{equation}
	loss_{ij} = max\left( {\frac{{t{p_i} - r{p_j}}}{{t{p_i}}},\frac{{t{p_j} - r{p_i}}}{{t{p_j}}}} \right)
\end{equation}

where ${tp}_\ast$ denotes the number of packets sent from port $\ast$, and ${rp}_\ast$ indicates the number of packets received from port $*$.

The LLDP and echo request with timestamps are used to measure the route delay \cite{RN35}. The controller sends echo requests to the switches s1 and s2. The delays between the controller and the switch ($d_{echo1}$ and $d_{echo2}$) can be calculated by minus the timestamp in the echo reply. And the delays between switches ($d_{lldp1}$ and $d_{lldp2}$) are calculated based on LLDP send and receive times. So the $delay_{ij}$ of  $e_{ij}$ is expressed in Formula \ref{eq:delay_ij}.
\begin{equation}
	delay_{ij} = \frac{{\left( {{d_{lldp1}} + {d_{lldp2}} - {d_{echo1}} - {d_{echo2}}} \right)}}{2}
	\label{eq:delay_ij}
\end{equation}

The current network link information (NLI) is calculated by the above methods and then saved to the \textit{network information storage} of the knowledge plane.

The control plane is also responsible for installing the flow table. If the packet fails to match the flow table, the switch sends a packet-in message to the controller \cite{RN5}. The controller then installs flow tables with forwarding information to the corresponding switches on the data plane via SBI based on the routing policy. When the switch is at the fork point of the multicast tree, the instructions in the flow entry set multiple apply actions to forward the flow to different ports.

\subsection{Knowledge plane}\label{subsec:Knowledge Plane}
The key work done in this paper is mainly concentrated here. The knowledge plane learns and decides the optimal multicast path based on the collected network link status information and sends the multicast tree to the control plane through the northbound interface.

First, the source node, multiple destination nodes and the NLIs stored in the \textit{network information storage} are processed into matrix form as the state of the DRL agent. The state matrix inputs into the neural network, and after training, the reward value converges, i.e., the agent learns the optimal strategy in the interaction with the environment, the optimal strategy $\pi^\ast$ is related to the remaining link bandwidth, link delay, and link packet loss rate in the reward function. We hope to construct a multicast tree based on $\pi^\ast$ with larger remaining bandwidth and a smaller delay and packet loss rate. Finally, the knowledge plane calculates the optimal multicast tree and submits to the controller, the routing decision corresponding to the multicast group is sent to the switches in the domain to implement the multicast routing. The detailed multicast routing algorithm based on deep reinforcement learning is introduced in Section \ref{sec:DRL-M4MR}.

\section{Design of DRL-M4MR intelligent SDN multicast routing algorithm}\label{sec:DRL-M4MR}
In this paper, the process of constructing a multicast tree is abstracted as a MDP. The MDP can be expressed as a quintuple $\left(\mathcal{S},\mathcal{A},P,R,\gamma\right)$, where $\mathcal{S}$ is the state set, $\mathcal{A}$ is the action set, $P$ is the state transition matrix, $R$ is the reward function, and $\gamma$ is the reward discount factor. At the current moment $t$, the agent observes the environment to obtain the current state $S_t\in\mathcal{S}$ and makes an action $A_t\in\mathcal{A}$ according to the strategy $\pi$ to obtain the corresponding reward $R_{t+1}$, and enters the next state $S_{t+1}$. However, the environment model, namely, the state transition function $P$, is unknown, so a model-free reinforcement learning method based on the value function is used in this paper. When the state space and the action space are large, the tabular reinforcement learning method is very inefficient, so deep neural networks are used to approximate the action-value function \cite{RN36}. In the state, $S_t$ inputs the neural network (policy network) and outputs the state-action value $Q\left(S_t,A_t\right)$ of different actions. The agent selects and executes action $A_t$ according to the $\varepsilon$-greedy. The target value is outputted from the target network according to the environment feedback reward $R_{t+1}$ and the next state $S_{t+1}$. The expected action value and the current action value calculate the loss, and the network parameter $\theta$ is updated by gradient descent. Then, the optimal strategy is found according to the next state $S_{t+1}$ iterative training.

\begin{figure}[ht]
	\centering
	\includegraphics[width=\linewidth]{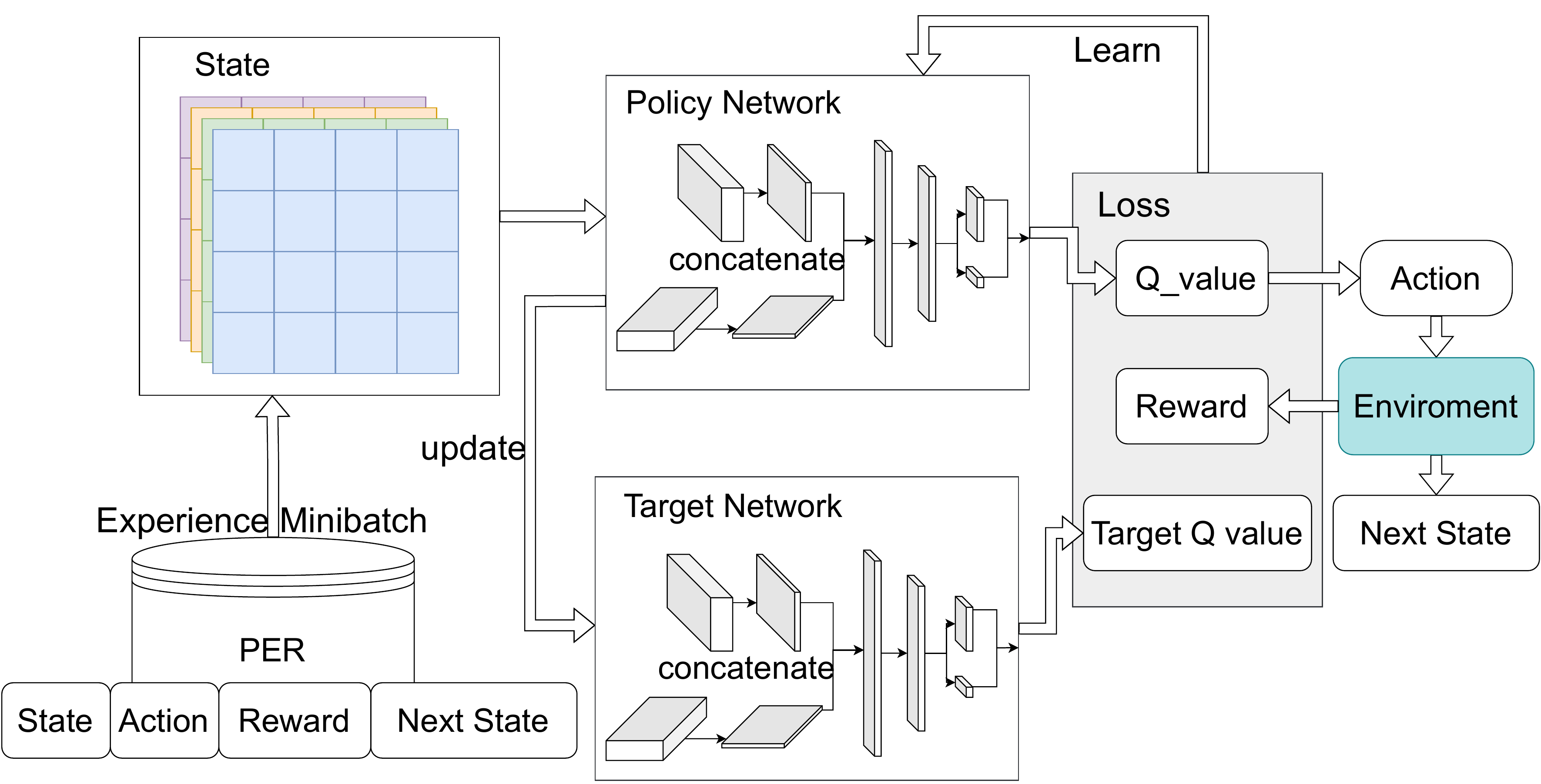}
	\caption{DRL-M4MR framework.}
	\label{fig:DRL_M4MR_framework}
\end{figure}

The design of the agent mainly includes the design of the state space, action space, and reward function. In this paper, improvement methods of deep reinforcement learning, such as decay $\varepsilon$-greedy and priority experience replay, are added to improve the learning effect and efficiency of the agent. The DRL-M4MR agent framework is shown in Figure \ref{fig:DRL_M4MR_framework}. First, the \textit{policy network} input is a minibatch of the state taken from PER, the output is the Q value, and action is selected according to the decay $\varepsilon$-greedy algorithm. The agent then interacts with the environment to produce a reward and the next state. The next state inputs into the target network and outputs the Q value. The value corresponding to the action is taken with the maximum Q value as the target value. Second, the parameters of the policy network are updated after the loss is calculated. Finally, the new state, action, reward, and next state are stored in the PER. The details of the algorithm design are as follows.

\subsection{State space}\label{subsec:state_space}
The state space is the current state that the DRL agent can observe through the environment. The current network in the data plane is regarded as a picture of four channels. The matrix of each channel contains the state information of the current multicast tree and multiple pieces of information for measurement on the data plane. The row number $i$ and column number $j$ of the matrix represent the node number. When $i\neq j$, the element of the $i$th row and $j$th column in the matrix represents the link $e_{\left(i,j\right)}$. And when $i$ is equal to $j$, the corresponding element represents node $i$, where $i,j\le n,e_{\left(i,j\right)}=e_{\left(j,i\right)}\in E$.
\begin{figure}[ht]
	\centering
	\begin{minipage}{0.45\linewidth}
		\subfigure[$M_T$ matrix]{
			\centering
			\includegraphics[width=0.8\linewidth]{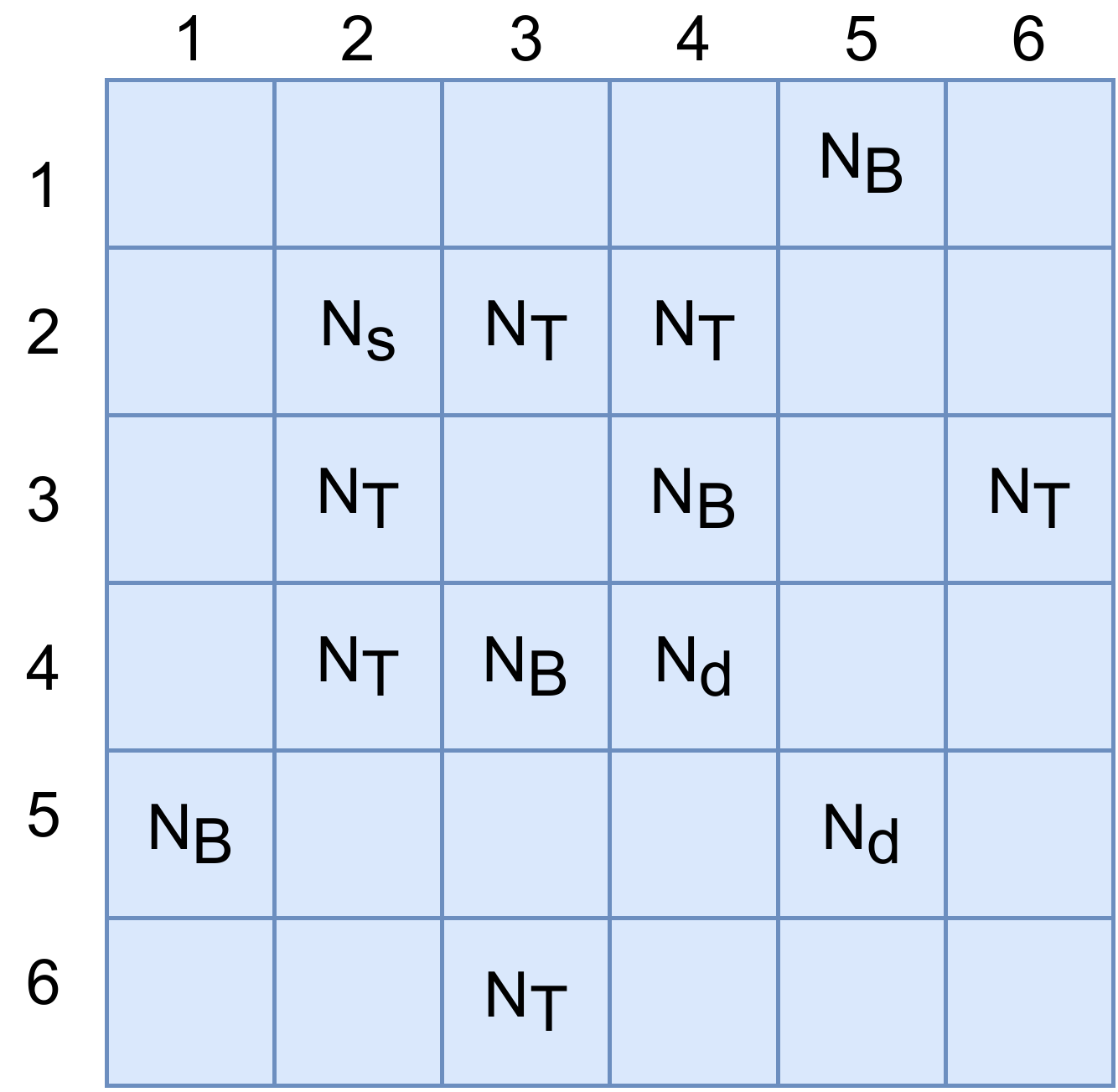}
			\label{subfig:Mt matrix}
		}
	\end{minipage}
	\begin{minipage}{0.45\linewidth}
		\subfigure[$M_B$,$M_{delay}$,$M_{loss}$ matrixes]{
			\centering
			\includegraphics[width=0.8\linewidth]{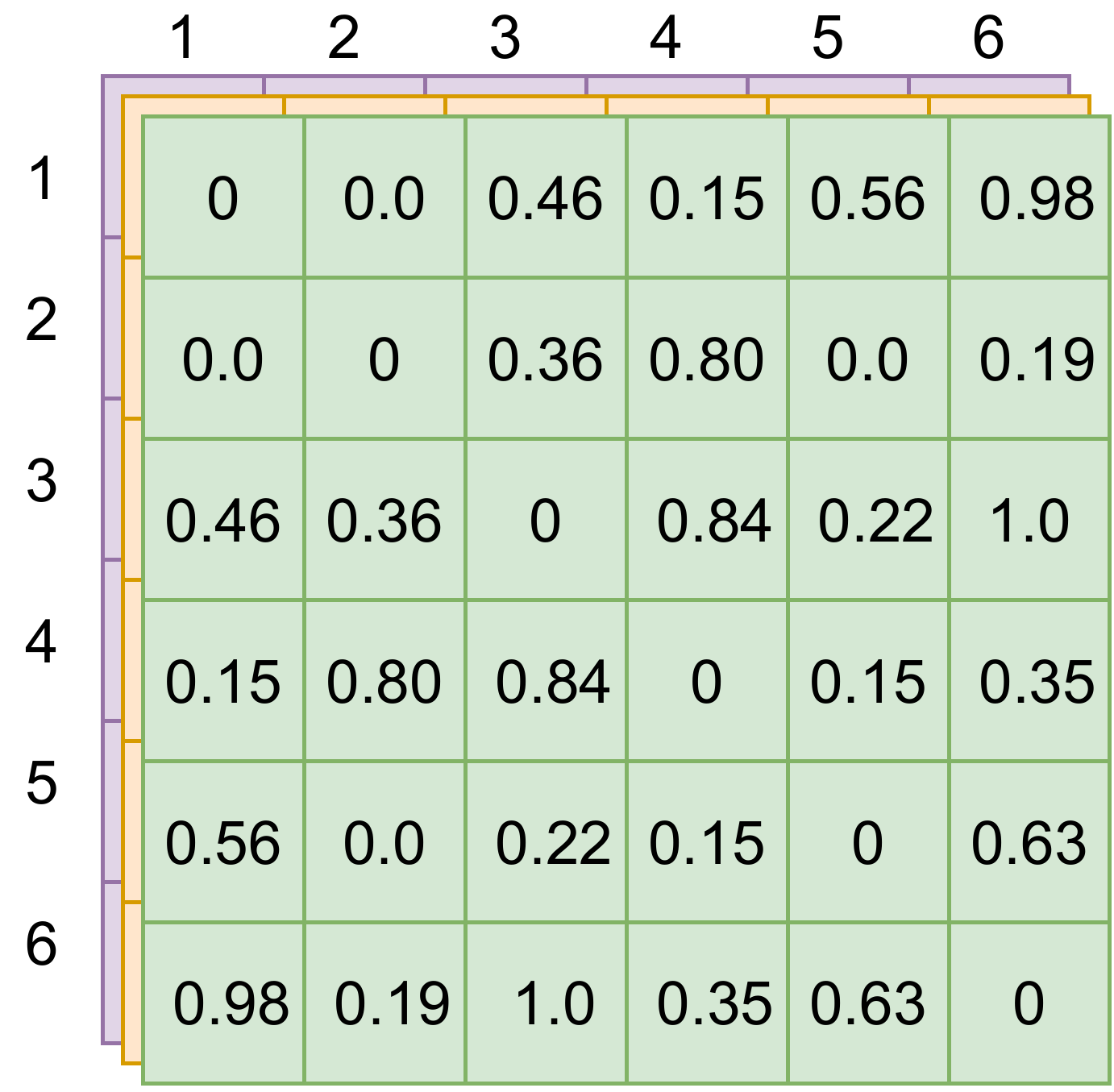}
			\label{subfig:Mb Mdelay Mloss matrixes}	
		}
	\end{minipage}
	\caption{The matrixes of state space.}
	\label{fig:state space matrix}
\end{figure}
The matrix $M_T$ of the first channel is a symmetric matrix that represents the state of the current multicast tree. The elements on the diagonal use the tag $N_s$ to represent the source node and the tag $N_d$ to represent the destination node. Moreover, the elements on the nondiagonal use the tag $N_T$ to represent the edge that is currently in the multicast tree. The tag $N_B$ represent the edges that can be added to the multicast tree and will not form a loop when added. After adding new edges, the matrix is modified as the next state $S_{t+1}$ according to the current tree. When all edges of the destination nodes are in the multicast tree, $S_{t+1}$ is set to terminate, i.e., $S_{t+1}=None$. As shown in Figure \ref{subfig:Mt matrix}, the source node is Node 2, the destination nodes are Node 4 and Node 5, and the edges that constitute the current multicast tree are $e_{\left(2,3\right)}$, $e_{\left(2,4\right)}$, $e_{\left(3,5\right)}$, and the edges that can be added are $e_{\left(3,4\right)}$, $e_{\left(1,5\right)}$.

The other three channel matrices represent the remaining link bandwidth matrix $M_{BW}$, the link delay matrix $M_{Delay}$, and the link packet loss rate matrix $M_{loss}$. The diagonal of the matrix is 0, and the remaining elements represent the normalized values of the measured link parameters in the network using Min-Max normalization, as shown in Figure \ref{subfig:Mb Mdelay Mloss matrixes}.

\subsection{Action space}
The action space is the set of actions that the agent can perform when interacting with the environment. The agent moves to the next state depending on the action. In the network topology with different degrees of nodes, the next-hop node cannot be used as the action space because the output of the neural network is a tensor of fixed shape. Inspired by the idea of the Prim algorithm \cite{RN37}, an edge that will not form a loop is added to the current tree at each iteration. 

We design all sets of edges as action spaces, $\mathcal{A}=\left\{e_1,e_2,\ldots,e_k\right\}$, where $e_k\in E$, $k=1,\ \ldots,m$. There are four cases to select an edge from $\mathcal{A}$ and join it to the multicast tree. (1) If the selected edge will not form a loop after being added to the tree (hereafter called optional branch), the added edge will be added to the multicast tree of $S_t$, then the multicast tree state matrix will be changed, and the agent will enter the next state $S_{t+1}$. (2) This edge will form a loop, so it is not added to the multicast tree, and the agent does not change the current state. (3) This edge is already in the tree, and the agent also does not change the current state. (4) This edge is not adjacent to the current tree and does not change the current state. 

\begin{figure}
	\centering
	\includegraphics[width=0.9\linewidth]{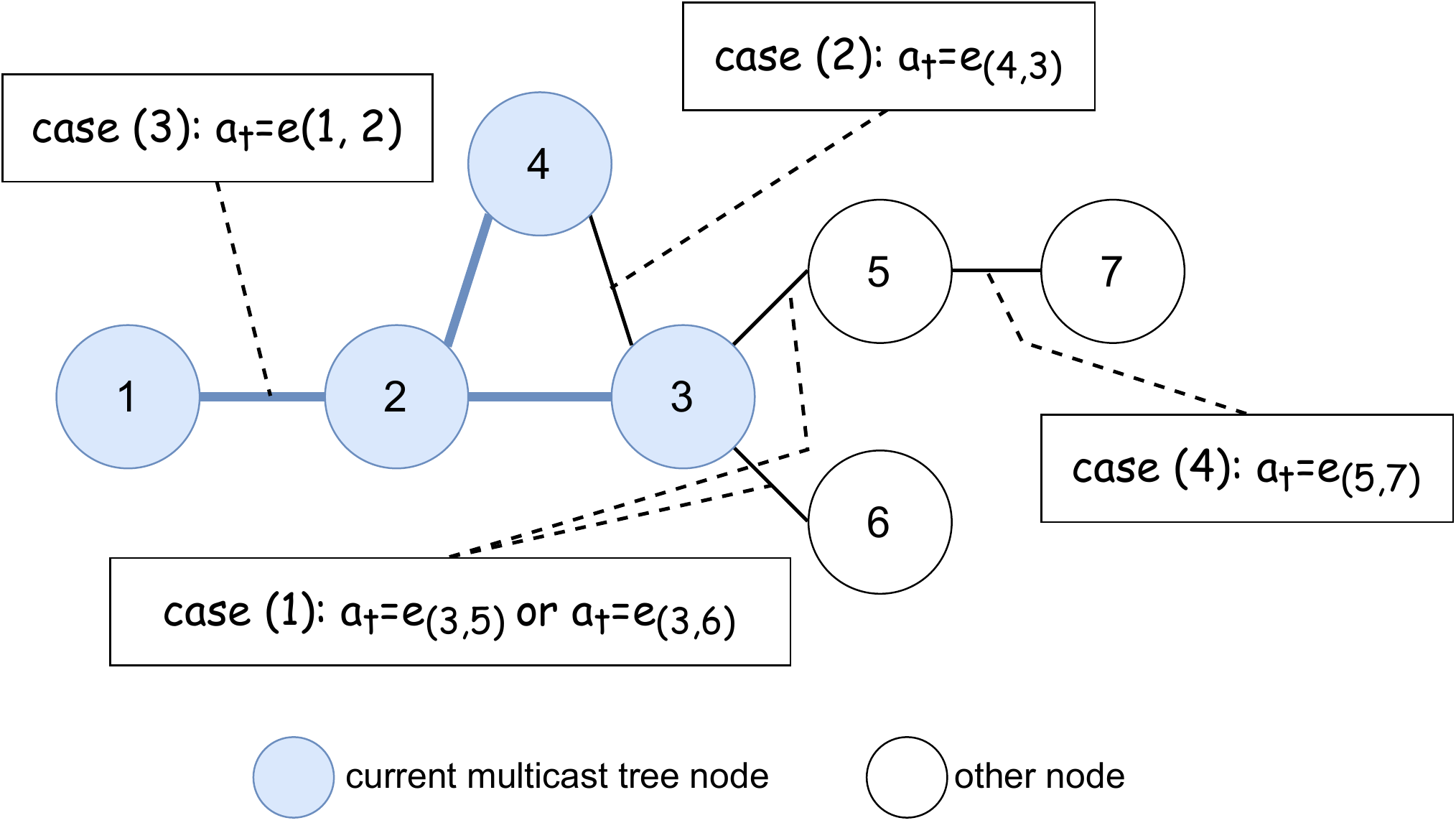}
	\caption{The four cases about action.}
	\label{fig:action_cases}
\end{figure}

As shown in Figure \ref{fig:action_cases}, blue nodes and edges represent the current multicast tree. If the action is $e_{\left(3,5\right)}$ or $e_{\left(3,6\right)}$, indicating Case (1), the edge can be added to the multicast tree. If the action is $e_{\left(4,3\right)}$, then Case (2) is indicated. If the action is $e_{\left(1,2\right)}$, Case (3) is indicated. If the action is $e_{(5,7)}$, as in Case (4), then, $e_{(5,7)}$ should not join the multicast tree.

\subsection{Reward function}
The reward value will guide the learning direction of the agent, and the agent will enter the ${t+1}$ after performing $A_t$ from the $S_t$. According to $A_t$ and $S_{t+1}$, the corresponding reward value $R_{t+1}$ can be obtained. Before the construction of the multicast tree is completed, the information of the whole tree cannot be obtained, so we can only evaluate the current decision from the link perspective. The agent can only be evaluated from a whole-tree perspective if it is in a final state. Three outcomes are generated when the agent interacts with the environment, and considering the optimized objective function, the reward function is designed as follows.

1) The new branch $e_{\left(i,j\right)}$ is added to the multicast tree, and the next state is not a complete state, that is, $S_{t+1}\neq None$, and the reward value is $R_{step}$ (Formula \ref{eq:R_step}). The agent is guided to choose the optimal link at each step, which is calculated according to the network link information, including the link available bandwidth $bw_{ij}$, the link delay $delay_{ij}$ and the link packet loss rate $loss_{ij}$. Min-Max normalization is performed for $bw_{ij}$, $delay_{ij}$ and $loss_{ij}$ to prevent one factor from having too much influence on decision-making.

\begin{equation}
	R_{step} = {\beta _1}b{w_{ij}} + {\beta _2}\left( {1 - dela{y_{ij}}} \right) + {\beta _3}\left( {1 - los{s_{ij}}} \right)
	\label{eq:R_step}
\end{equation}
where $\beta_1$, $\beta_2$ and $\beta_3\in\left[0,1\right]$ are tunable parameters to provide a weight value for the reward calculation.

2) If the edge selected by the policy cannot be added to the current tree or is already in the multicast tree, a constant value penalty is applied, $R_{trap}=C$.

3) When a new branch is added to the multicast tree, $S_{t+1}$ becomes the final state, indicating that the agent completes the construction of the multicast tree from the source node s to the destination node set D. After that, the reward value $R_{finish}$ (Formula \ref{eq:R_finish}) is calculated from the four factors of bandwidth, delay, packet loss rate, and redundant branches of the whole tree.
\begin{equation}\label{eq:R_finish}
	{R_{finish}} = {\beta _1}b{w_{tree}} + {\beta _2}\left( {1 - dela{y_{tree}}} \right)\\ 
	+ {\beta _3}(1 - los{s_{tree}})
\end{equation}

The reward value $R_{finish}$ is proportional to the $bw_{tree}$ and inversely proportional to the $delay_{tree}$ and $loss_{tree}$. $\beta_1$, $\beta_2$ and $\beta_3$ are defined the same as Formula \ref{eq:R_step}. Additionally, to prevent one factor from influencing the reward value too much, the $bw_{tree}$ and $delay_{tree}$ are reduced to between 0 and 1 according to Formula \ref{eq:bw_tree_hat} and Formula \ref{eq:delay_tree_hat}.
\begin{equation}\label{eq:bw_tree_hat}
	{\widehat {bw}_{tree}} = \frac{{bw_{tree} - min(bw_{ij})}}{{max(bw_{ij}) - min(bw_{ij})}}
\end{equation}
\begin{equation}\label{eq:delay_tree_hat}
	{\widehat {delay}_{tree}} = \frac{{dela{y_{tree}} - min(dela{y_{ij}}) \cdot |S|}}{{\sum\limits_{{e_{ij}} \in E} d ela{y_{ij}} - min(dela{y_{ij}}) \cdot |S|}}
\end{equation}
where $\theta_v$ and $\theta_a$ denote the parameters of the fully connected layer of the state-value function and advantage function in the \textit{policy network}, respectively.

\subsection{Q value function}
The action-value function $Q\left(s,a\right)$ evaluates the value of action $a$ performed by an agent in state $s$, updates the current $Q\left(s,a\right)$ from the next observed state, and returns by temporal difference.

To reduce the overestimation problem of a single network and improve the stability of the Q value, the double dueling DQN \cite{RN38} method is adopted. The double DQN calculates a temporal difference (TD) error between the current Q value and target value, as Formula \ref{eq:TDerror}. Then, the parameter $\theta$ is updated by gradient descent, where $\theta$ and $\hat{\theta}$ assume the parameters of two deep neural network. where $\theta$ and $\hat{\theta}$ denote the parameters of the \textit{policy network} and \textit{target network}, respectively.
\begin{equation}\label{eq:TDerror}
\begin{split}
TD_{error} = &{R_{t+1}}+\gamma \hat Q\left( {{S_{t+1}},argmax_{{a^\prime }}\hat Q\left( {{S_t},{a^\prime };\hat \theta } \right);\hat \theta } \right)\\
&-Q\left( {{S_t},{A_t};\theta } \right)
\end{split}
\end{equation}

The dueling DQN decoupled the Q value into the state value function $V$ and action advantage function $A$. $V$ and $A$ evaluate the current state and the importance of each action, respectively, making the training more stable.
\begin{equation}\label{eq:dueling Q}
\begin{split}
&Q\left( {{S_t},{A_t};\theta ,{\theta _v},{\theta _a}} \right) =V\left( {{S_t};\theta ,{\theta _v}} \right)\\
&+ \left( {A\left( {{S_t},{A_t};\theta ,{\theta _a}} \right) - \frac{1}{{\left| {\cal A} \right|}}\sum\limits_{{a^\prime }} {A\left( {{S_t},{a^\prime };\theta ,{\theta _a}} \right)} } \right)
\end{split}
\end{equation}
where $\theta_v$ and $\theta_a$ denote the parameters of the fully connected layer of the state-value function and advantage function in the \textit{policy network}, respectively.

\subsection{Exploration method}
In the early stages of training, the agents are expected to explore more possible results, while there is good convergence in the later stages. The decay $\varepsilon$-greedy (Formula \ref{eq:egreedy}) is used to select actions, $x\sim U\left(0,1\right)$. When $x\geq \varepsilon$, the largest value function of the current action is utilized; otherwise, an action is randomly selected in the action space.
\begin{equation}\label{eq:egreedy}
	{a_t} = \left\{ {\begin{array}{*{20}{c}}
		{argmax_aQ\left( {{s_t},a} \right),\quad if\quad x \ge {\varepsilon}}\\
		{random \quad choice,\quad otherwise}
		\end{array}} \right.
\end{equation}
where $\varepsilon$ is calculated by Formula \ref{eq:epsilon}.
\begin{equation}\label{eq:epsilon}
	{\varepsilon} = {{\varepsilon}_{final}} + \left( {{{\varepsilon}_{start}} - {{\varepsilon}_{final}}} \right) \cdot exp\left( { - \frac{{episode_{curr}}}{{episode_{decay}}}} \right)
\end{equation}
where $\varepsilon_{start}$ denotes the value of $\varepsilon$ at the beginning of training. The $episode_{curr}$ is the current training episode. The $episode_{decay}$ is used to adjust the episode of convergence, as the training continues and $\varepsilon$ tends to $\varepsilon_{final}$, and $\varepsilon_{start}$, $\varepsilon_{final}\in\left[0,1\right]$.

\subsection{Prioritized experience replay}
The agent will produce the transition $\langle S_t,A_t,R_{t+1},S_{t+1}\rangle$ as the learning experience is stored in the experience pool at every decision. The PER \cite{RN39} assigns different sampling priorities to each transition according to its TD error; the larger the TD error is, the more important the transition will be. PER selects effective actions as learning experiences, which can improve the learning efficiency of agents.

The sampling probability is proportional to the TD error, $i\sim P\left(i\right)\propto\left|TD_{error}\right|^\alpha$, where $\alpha$ is a hyperparameter that determines the shape of the distribution. Finally, the sampling parameter $\omega_i$ is important because it is used to correct the deviation of the distribution.

\subsection{Multi-step learning}
Traditional DQN agents estimate the target value through a one-step reward and the $S_{t+1}$ value. When the decision at $S_t$ produces a deviation, it will lead to a deviation in future decisions. Therefore, when calculating the current return $R_{t+1}$, considering the future n steps return $R_{t:t+n}$ (Formula \ref{eq:R_multistep}) will make the estimation more accurate and can speed up the learning speed \cite{RN40}.
\begin{equation}\label{eq:R_multistep}
	{R_{t:t + n}} = \sum\limits_{k = 0}^{n - 1} {{\gamma ^k}{R_{t + 1 + k}}}
\end{equation}
The TD(n) error needs to be calculated after using multi-step learning:
\begin{equation}
\begin{split}
&T{D_{error}}\left( n \right) ={R_{t:t + n}}\\
&+ \gamma ma{x_a}Q\left( {{S_{t + n}},a;\hat \theta } \right) - Q\left( {{S_t},{A_t};\theta } \right)
\end{split}
\end{equation}

\subsection{DRL-M4MR multicast routing algorithm based on DQN}
According to the input source node $s$ and destination node-set $D$, DRL-M4MR finds the optimal multicast tree from $s$ to $D$ in the currently observed network environment topology $G$. The agent follows Algorithm \ref{alg:DRL-M4MR} during training and needs the following inputs: the learning rate $\alpha$; the batch-size $k$ of each collection from the PER pool; the updated frequency $n_{update}$, which indicates that the parameters of the policy network are updated to the target network at each $n_{update}$ step during training; and the total number of  training episodes $M$. The output is the multicast tree from $s$ to $D$ in $G$.

Line 1 to Line 3 are initialization. The parameter $\theta$ of the policy network is randomly initialized, and this parameter is copied to initialize the target network parameter $\hat{\theta}$. The PER pool is initialized to empty. In Line 6, the agent initializes the multicast state matrix $M_T$ based on the inputs $N_s$ and $N_d$, and the current optional branch of the tree is the link connected to the initial node. In Line 7, $M_T$, $M_{bw}$, $M_{delay}$ and $M_{loss}$ are connected in the channel dimension, and the matrix of dimensions $(4×n×n)$ is taken as the initial state $S_t$. Then, the training begins in the environment of different residual bandwidth, delay, and packet loss rates. 

Line 4 to Line 28 conduct the training of an episode. After adding all destination nodes to the multicast tree, the agent considers that it has completed the training of the current environment and observes the initialization state of the environment $S_t$ for new training when the data stored in the network information storage are traversed, and the training of the episode will be completed.

\begin{algorithm}[ht]\small
	\label{alg:DRL-M4MR}
	\caption{DRL-M4MR}
	\LinesNumbered 
	\KwIn{source node $N_s$,
		destination nodes set $N_d$,
		learning rate $\alpha$,
		n-step $n$,
		batch-size $k$,
		soft-update frequency $n_{update}$,
		training episodes $M$.
	}
	\KwOut{optimal multicast tree for $(N_s, N_d)$.}
	\BlankLine
	Initialize policy network with random weight $\theta$\;
	Initialize target network with weight $\hat{\theta}=\theta$\;
	Initialize PER pool $PER$, n-steps buffer $B$\;
	\For{$episode=1\leftarrow M$}{
		\For{$M_{bw},M_{delay},M_{loss}$ in Network Information Storage}{
			Reset enviroment with $(N_s, N_d)$ \;
			Get $s_t \leftarrow cat(M_T,M_{bw},M_{delay},M_{loss})$\;
			\tcp{concatenate in the channel dimension}
			\While{True}{
				Take an action $a_t$ from $s_t$ according to decay $\varepsilon$-greedy\;
				Execute action $a_t$ and observe reward $r_t$ and next state $s_{t+1}$\;
				Store $\langle s_t, a_t, r_{t}, s_{t+1}\rangle$ in $B$\;
				\If{$len(PER)\geq k$}{
					Sample minibatch $\langle s_i, a_i, r_{i:i+n}, s_{i+n}\rangle$ and get $\omega_i$ from $PER$\;
					\tcp{$\omega_i$ is importance-smapling weight}
					Get $Q(s_i,a_i)$ from policy network\;
					\lIf{$s_{i+n}$ is not None}{
						$G\leftarrow R_{i:i+n} + \gamma max_{a'}\hat{Q}(s_{i+n},a')$ }
					\lElse{
						$G\leftarrow R_{i:i+n}$
					}
					
					Compute TD-error $\delta_i\leftarrow G-Q(s_i,a_i)$\;
					Update PER transition priority by $\delta_i$\;
					Update policy network paramters $\theta \leftarrow \theta + \alpha\omega_i\delta_i\nabla_\theta Q(s_i, a_i)$ \;
					
				}
				\If(\tcp*[f]{all destination nodes in multicast tree}){done}{
					Break\;} 
				$s_t \leftarrow s_{t+1}$\;
				In every $n_{update}$ steps update, set $\hat\theta\leftarrow \theta$\;
			}
		}
	}
	Use final policy network with parameter $\theta$ to construct multicast tree from $N_s$ to $N_d$, 
	the agent execute the max Q-value action for each decision.  
\end{algorithm}

From Line 9 to Line 11, the agent selects action $a_t$ in the current state $s_t$ according to the decay $\varepsilon$-greedy, executes $a_t$ and returns $r_t$ from the environment and observes the next state $s_{t+1}$, storing the transition in the PER pool.

From Line 12 to Line 19, a minibatch of size k is sampled from the PER if the number of stores in the PER is greater than or equal to the batch size. The $Q\left(s_i,a_i\right)$ and $max_{a^\prime}\hat{Q}\left(s_{i+1},a^\prime\right)$ are calculated from the policy network and target network, respectively, where $i\le k$. If $s_{i+1}$ is not a final state, then according to the action-value function, the target reward $R\prime$ is calculated; otherwise, $R\prime$ is the final reward value. Then, the priority in the PER is updated according to the calculated TD error $\delta_i$. The policy network parameters are updated using gradient descent and backpropagation algorithms to minimize the TD error. From Line 21 to Line 23, the loop is broken if it is in the final state; otherwise, $s_t$ is set equal to $s_{t+1}$, and the next action selection starts until all destination nodes are added to the multicast tree. After each $n_{update}$ generation training, the agent can use $\theta$ to update the \textit{target network} parameter $\hat{\theta}$ bearing (Line 25).

Using the parameters completed by the policy network training, the agent selects the action with the maximum value of the action state in each state as the decision of each step to select the multicast tree from $N_s$ to $N_d$ (Line 29). After the training, because an edge is selected to be added to the multicast tree at each step of decision-making, the complexity in the worst case is $O(N)$, where $N$ is the number of edges in $G$.

After training through Algorithm 1, the agent uses Algorithm \ref{alg:remove_redundancy} to construct a multicast tree and deliver flow entries. When constructing a multicast tree (Line 5 to Line 13), the parent of the current node is stored in the dictionary $agent\_route\_dict$ by unpacking the link corresponding to an action $A_t$ into two nodes.

\begin{algorithm}[ht]\small
	\label{alg:remove_redundancy}
	\caption{Install multicast flow entry}
	\LinesNumbered 
	\KwIn{DRL-M4MR agent with parameter $\theta$,
		source node $N_s$,
		destination nodes set $N_d$,
		current network information $M_{BW}$,$M_{delay}$,$M_{loss}$.
	}
	\BlankLine
	Initialize agent\_route\_dict=\{$N_s$: None\} \;
	Initialize install\_info\_dict=\{$N_s$:\{'parent': None, 'child': []\}\} \;
	Reset enviroment with $(N_s, N_d)$ \;
	Get $s_t \leftarrow cat(M_T,M_{BW},M_{delay},M_{loss})$\;
	
	\While(\tcp*[f]{Contruct multicast tree}){True}{
		Take a max $Q$ value action $a_t$ from $s_t$ \;
		Execute action $a_t$ and observe next state $s_{t+1}$ \;
		parent\_node, current\_node=$a_t$ \;
		agent\_route\_dict[current\_node]=parent\_node \;
		
		\If{all destination nodes in multicast tree}{Break\;} 
		$s_t=s_{t+1}$ \;
	}

	\For(\tcp*[f]{Delete redundant node}){node in $N_d$}{
		\While{node is not None}{
			parent\_node =agent\_route\_dict[node] \;
			install\_info\_dict[node]['parent'] = parent\_node \;
			install\_info\_dict[parent\_node]['child'].append(node) \;
			
			node $\leftarrow$ parent\_node \;
			\If{node in install\_info\_dict}{
				\tcp{node have been travesed}
				Break\;}
		}
	}
	
	\For(\tcp*[f]{Install flow entry}){node in install\_info\_dict}{
		parent\_node = install\_info\_dict[node]['parent']\;
		child\_node = install\_info\_dict[node]['child']\;
		in\_port, out\_ports $\leftarrow$ get\_ports(node, parent\_node, child\_node)\;
		Install multicast flow to node corresponding switch \;
	}
	
\end{algorithm}

Redundant branches are generated in the construction of the multicast tree. To ignore the redundant branches, only the reverse traverse from the leaf node is the destination node. In a tree structure, there is a unique path from the root node to any leaf node, so there is a unique path from the leaf node to the root node. If the leaf node is not the destination node, the node is redundant, and the subtree in which it is located in a redundant branch. If the root of the subtree is only one subtree, the subtree where the root resides is also a redundant branch. When deleting redundant branches (Line 15 to Line 25), the parent node is queried at the $agent\_route\_dict$ and traversed in reverse from each destination node to the root node. If the node is already traversed, the parent and child nodes of the current node are stored in $install\_Info\_dict$.

Finally, Algorithm \ref{alg:remove_redundancy} traverses $install\_Info\_dict$ (Line 26 to Line 31), obtains the corresponding inbound and outbound ports according to the parent node and child node, and delivers the flow table entry that forwards data from the inbound port to the outbound port to the switch.

\section{Experiments and evaluation}\label{sec:experiments}
The control plane of the experiment uses the SDN controller Ryu\cite{ryu.org}. The data plane uses Mininet 2.3.0 \cite{mininet.org} to simulate the network. It is deployed on the Ubuntu 20.04.3 server that runs the GeForce RTX 2080 graphics card. Ryu provides the southbound interface OpenFlow 1.3 to communicate with Mininet's Open vSwitches. Python scripts are utilized to collect network information and store it as Comma-Separated Values (CSV) files and pickled files. Reinforcement learning and deep neural networks are implemented using Python 3.8 and PyTorch 1.9.0.

\begin{figure}[h]
	\centering
	\includegraphics[width=\linewidth]{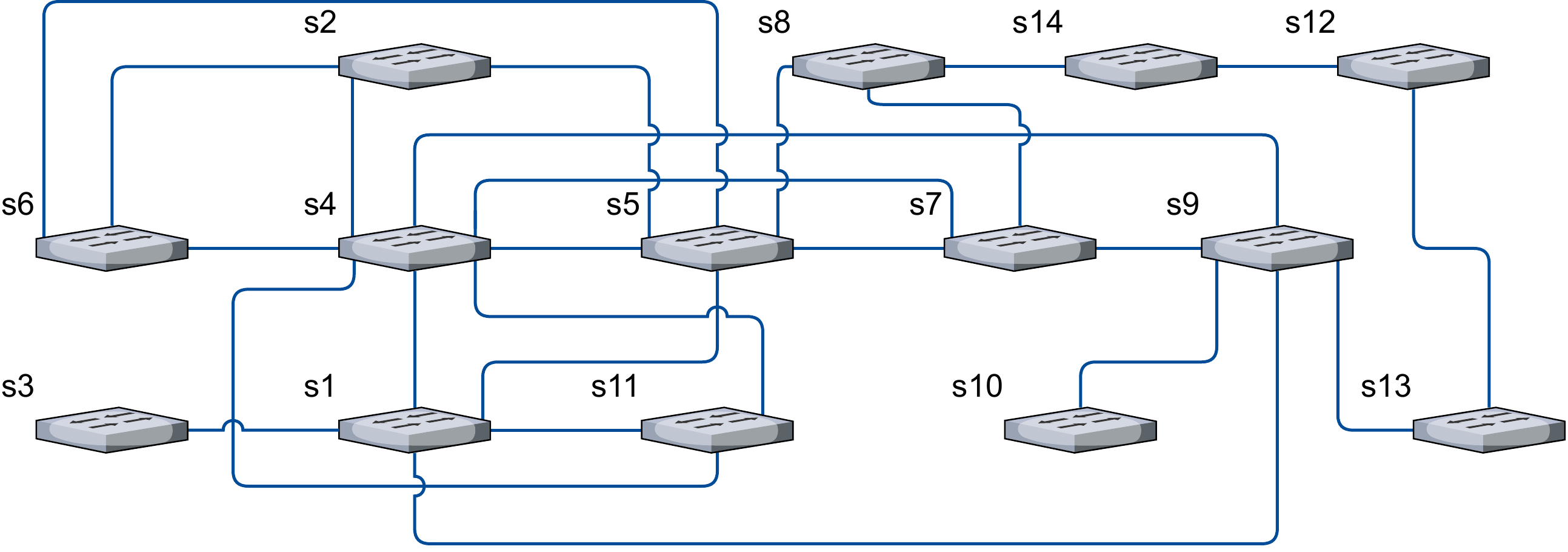}
	\caption{The experimental topology.}
	\label{fig:topo}
\end{figure}

The experimental topology Figure \ref{fig:topo} shows the edge data center topology in New York City \cite{topo.org}, which contains 14 nodes and 23 links. The link parameters are randomly generated, and the bandwidth and delay are set in the range of 5$\sim$30 Mbps and 1$\sim$20 ms, respectively, which obey a uniform distribution.

\begin{figure}[h]
	\centering
	\includegraphics[width=0.8\linewidth]{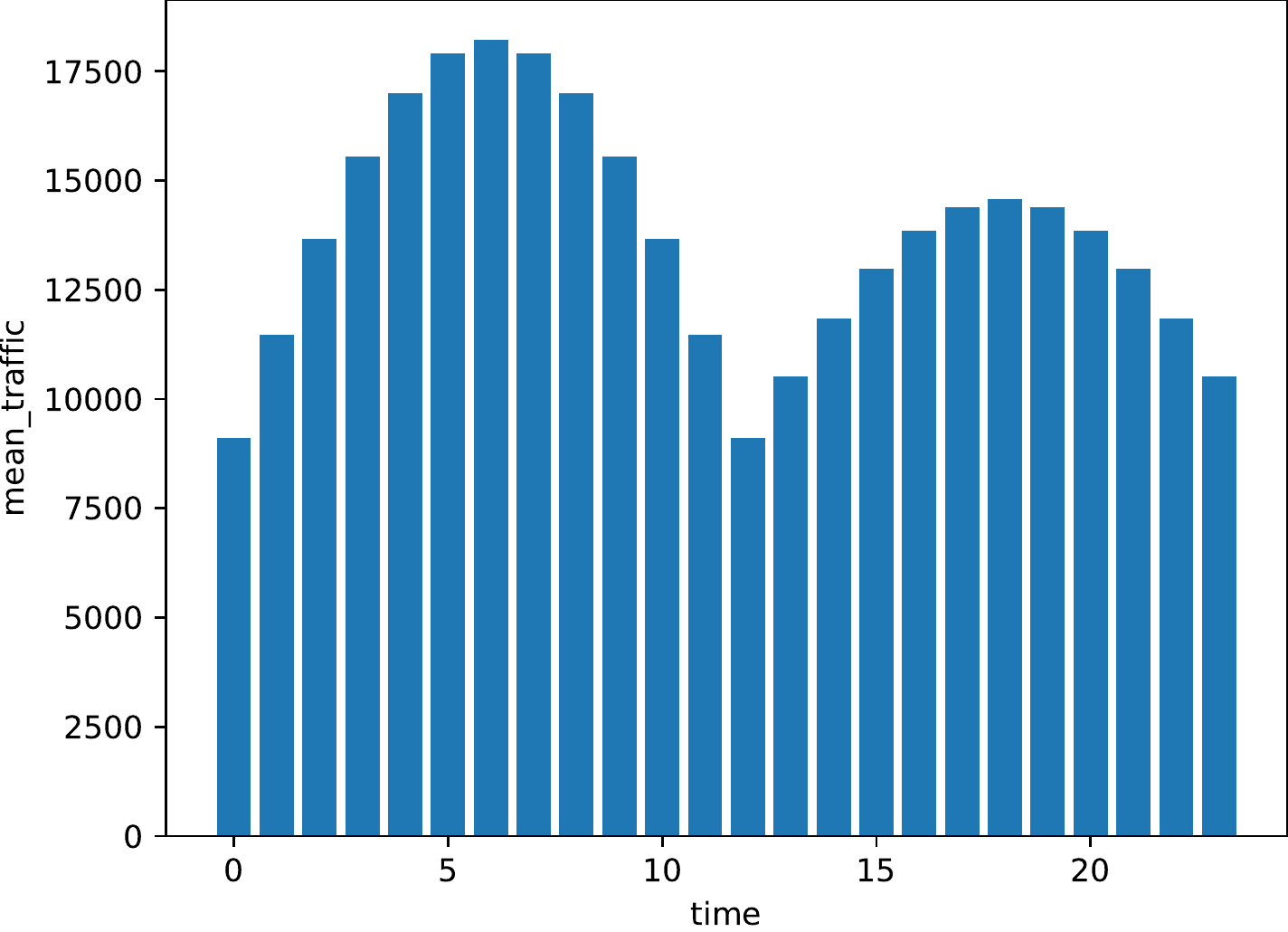}
	\caption{Mean bandwidth of the traffic matrix.}
	\label{fig:tm}
\end{figure}

To simulate network traffic, we use the traffic matrix generation tool \cite{RN43} to generate 24 traffic matrices between 14 nodes. Each element of the matrix represents the amount of traffic sent from the source node i\ to the destination node j. As shown in Figure \ref{fig:tm}, the ordinate represents the average amount of traffic sent by each node in Kbits/sec. In Mininet, each node is regarded as a server node and a client nod. The Iperf \cite{iperf.org} tool is used to send UDP packets with specified data sizes between different nodes to simulate traffic and achieve dynamic changes in the network environment. Finally, the Ryu controller measures the network parameters and stores the measurement results.

\subsection{DRL paramters setting}
The input state matrix passes through two convolution layers with two convolution kernels, which extract features using convolution kernels of size 5×1 and 1×5 After flattening the features into a one-dimensional tensor, they are combined in the element dimension through a concatenation operation. After the combination, two fully connected layers are input. The output of the fully connected layer is inputted into the fully connected layer of the advantage function and the fully connected layer of the value function. The policy network and target network are shown in Figure \ref{fig:DRL_M4MR_framework}. Finally, the output of the Q value is calculated according to Formula \ref{eq:dueling Q}. The LeakyReLU activation function is used after each convolution layer and full connection layer. The optimizer uses adaptive moment estimation (Adam), which has the advantages of an adaptive learning rate and momentum gradient to prevent falling into local solutions and helps the neural network to better converge.

This part of the experiment is tested under three network status information scenarios, that is, the controller collected three network status information and converted it into three groups of $M_{BW}$, $M_{delay}$ and $M_{loss}$. Because it is easy to obtain the optimal multicast tree under a small number of scenarios, we can better observe whether the convergence of the agents is optimal. The source node is 12, and the destination node-set is $\{2,4,11\}$ because the multicast tree from the source node to the destination node is expected to be not unique in the experiment, and the agent is expected to face more link choices when constructing the multicast tree to test the effectiveness of the algorithm.

The reward settings include $R_{step}$, $R_{trap}$, and $R_{finish}$. Agents tend to obtain larger reward values, and different ratios of reward and penalty values will affect the convergence and even lead to nonconvergence.

We set $R_{trap}=-1$, and experiments show that when the upper limit for $R_{finish}$ and $R_{step}$ is set to 1, $R_{finish}:R_{step}=1:1$, the reward value of each step decision is as important as the final reward value. In this case, the agent will choose the link with the higher reward value in the first few steps of the state so that the network parameters will receive many updates, and thus the Q value will be updated, although this decision will lead to a redundancy link in the future.

When the agent reaches the final state $S_{t+k}$ and selects an action, a link with a higher reward value than that of reaching the final state can be selected, and the agent chooses this link in preference, resulting in the construction of the multicast tree containing other redundant branches.

Setting $R_{finish}:R_{step}=1:0.1$ causes both the reward value of each step decision and the final reward value to take effect. $R_{step}$ guides the agent to choose the current optimal link, and $R_{finish}$ evaluates and awards the agent from the perspective of the whole tree. The agent will seek the optimal parameter of the whole tree. However, fewer redundant links are generated for the same reasons described in the previous section.

Setting $R_{finish}:R_{step}=1:0.01$ causes the reward value of each step decision to be far less than the final reward value. Thus, the update of the network parameters mainly depends on the Q value of the final decision, so that the Q value of the state is closer to the final state $S_{t+k}$ updates more easily, and the intelligence chooses the shortest path to reach the destination node possible. However, the bandwidth, delay, and packet loss rate of this tree are not optimal.

We also tried different reward value designs, which are summarized in Table \ref{tab:rewardslist}.

\begin{table}[h]\small
	\caption{reward value setting}
	\begin{tabular}{|c|c|c|c|c|c|} \hline
		$R_{finish}:R_{step}$ & $bw_{tree}$ & $delay_{tree}$ & $loss_{tree}$ & steps & redundancy\\ \hline
		1:1 & 13.69 & 13.31 & 3.69e-4 & 8.0 & 5.0\\ \hline
		1:0.1 & 14.35 & 13.22 & 3.69e-4 & 7.5 & 3.5\\ \hline
		1:0.01 & 14.04 & 13.07 & 4.91e-3 & 7.0 & 0.0\\ \hline
		1:-0.01 & 14.04 & 13.07 & 4.91e-3 & 7.0 & 0.0\\ \hline
		1:-0.1 & 14.04 & 13.07 & 4.91e-3 & 7.0 & 0.0\\ \hline
		1:-1 & 14.04 & 13.07 & 4.91e-3 & 7.0 & 0.0\\ \hline
	\end{tabular}
	\label{tab:rewardslist}
\end{table}
Redundancy indicates the average number of redundant links. The $bw_{tree}$, ${delay}_{tree}$, and $loss_{tree}$ are calculated according to Formula \ref{eq:bw_tree}, Formula \ref{eq:delay_tree}, and Formula \ref{eq:loss_tree}, respectively, and take the average value after calculation in different scenarios. Steps indicate the average number of steps an agent takes to decide on constructing a multicast tree. When the single-step reward value is negative, the value is calculated according to Formula \ref{eq:neg_R_step}.
\begin{equation}\label{eq:neg_R_step}
	{R_{step}} =  - ({\beta _1}(1 - b{w_{ij}}) + {\beta _2}dela{y_{ij}} + {\beta _3}los{s_{ij}})
\end{equation}

\begin{figure*}[ht]
	\centering
	\begin{minipage}{0.3\textwidth}
		\subfigure[lr]{
			\centering
			\includegraphics[width=1\textwidth]{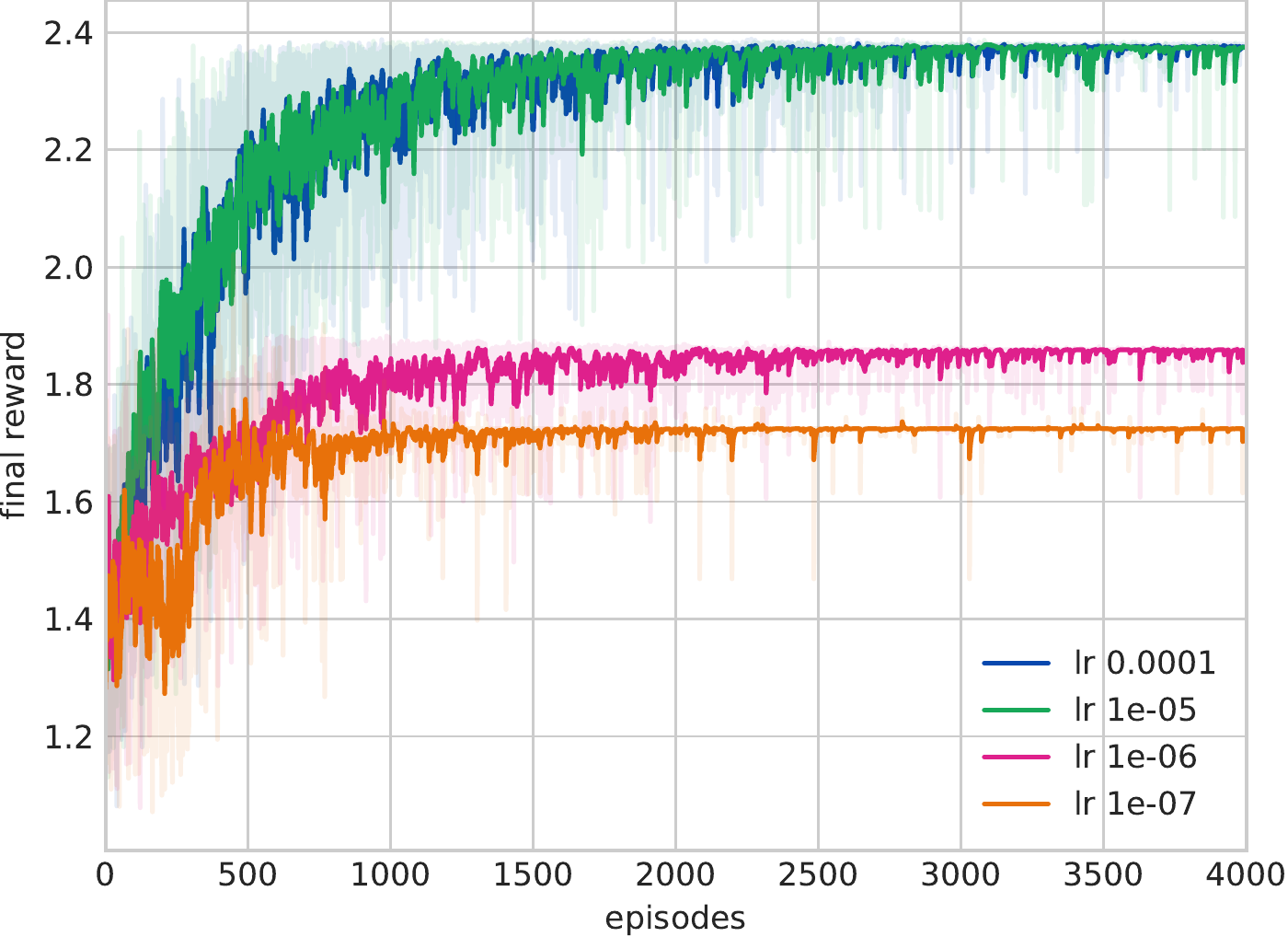}\hspace{-15pt}
			\label{subfig:lr}
		}
	\end{minipage}
	\begin{minipage}{0.3\textwidth}
		\subfigure[batchsize]{
			\centering
			\includegraphics[width=1\textwidth]{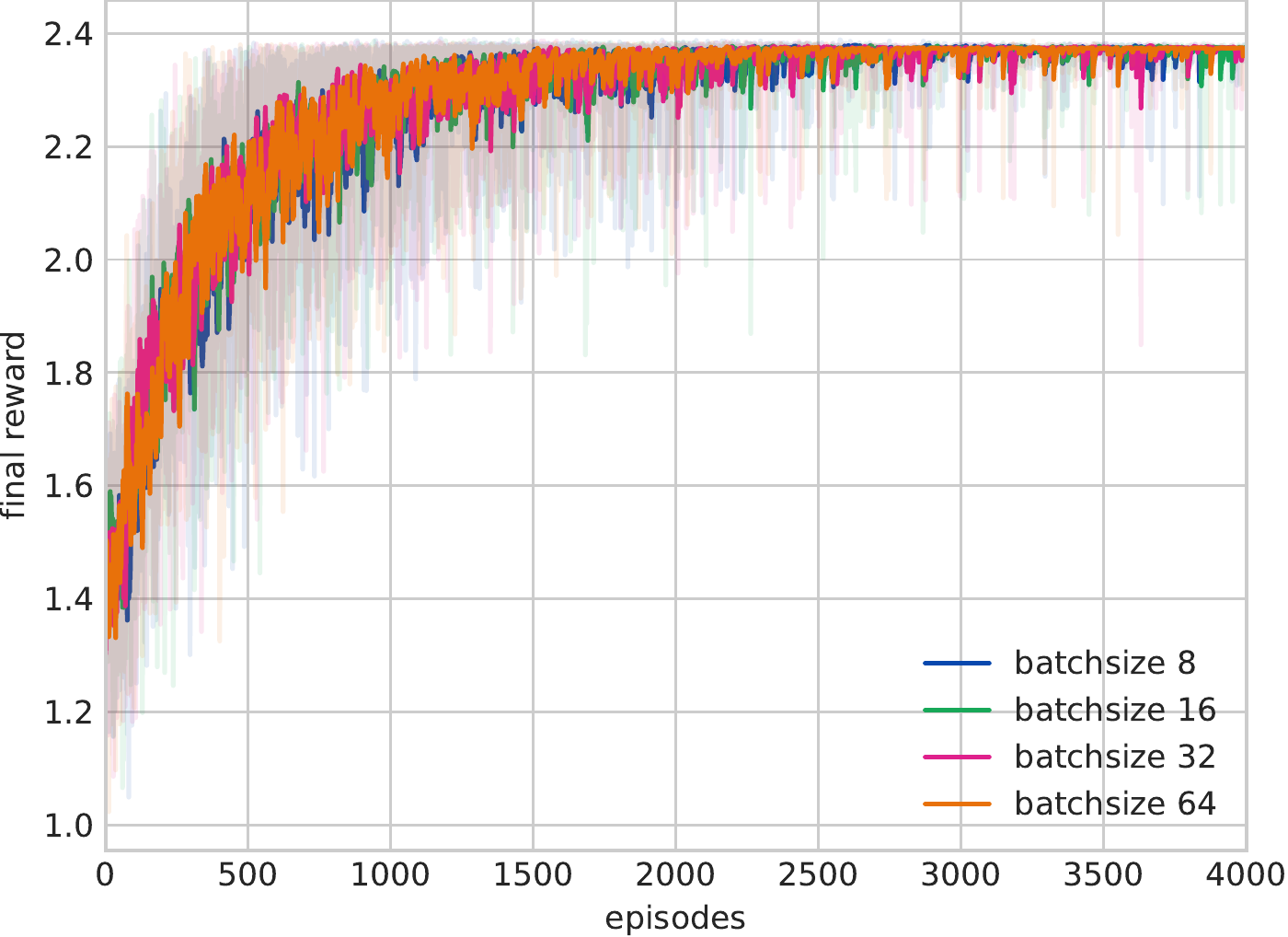}\hspace{-15pt}
			\label{subfig:batchsize}
		}
	\end{minipage}
	\begin{minipage}{0.3\textwidth}
		\subfigure[nsteps]{
			\centering
			\includegraphics[width=1\textwidth]{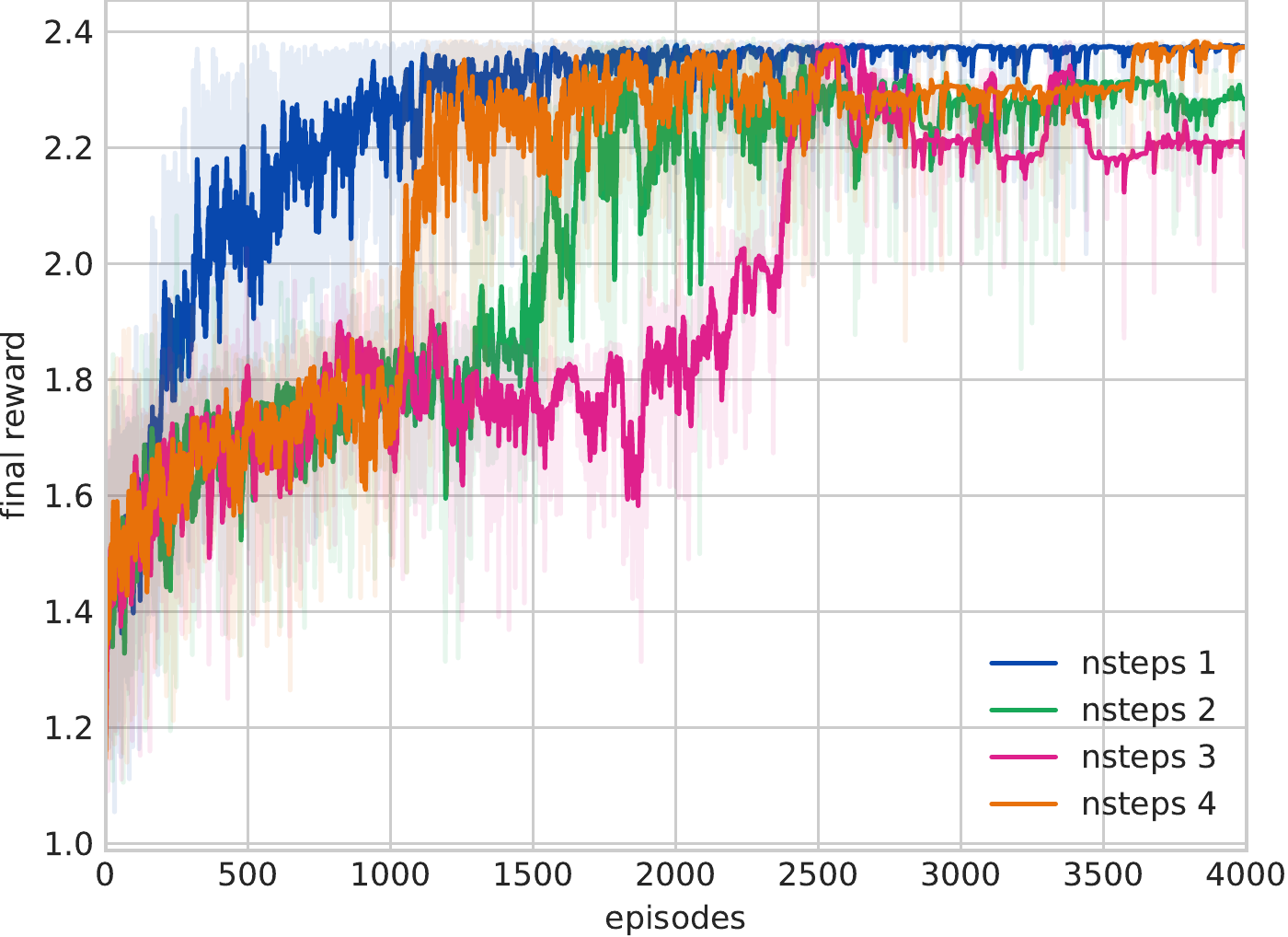}\hspace{-15pt}
			\label{subfig:nsteps}
		}
	\end{minipage}
	\begin{minipage}{0.3\textwidth}
		\subfigure[updatefrequency]{
			\centering
			\includegraphics[width=1\textwidth]{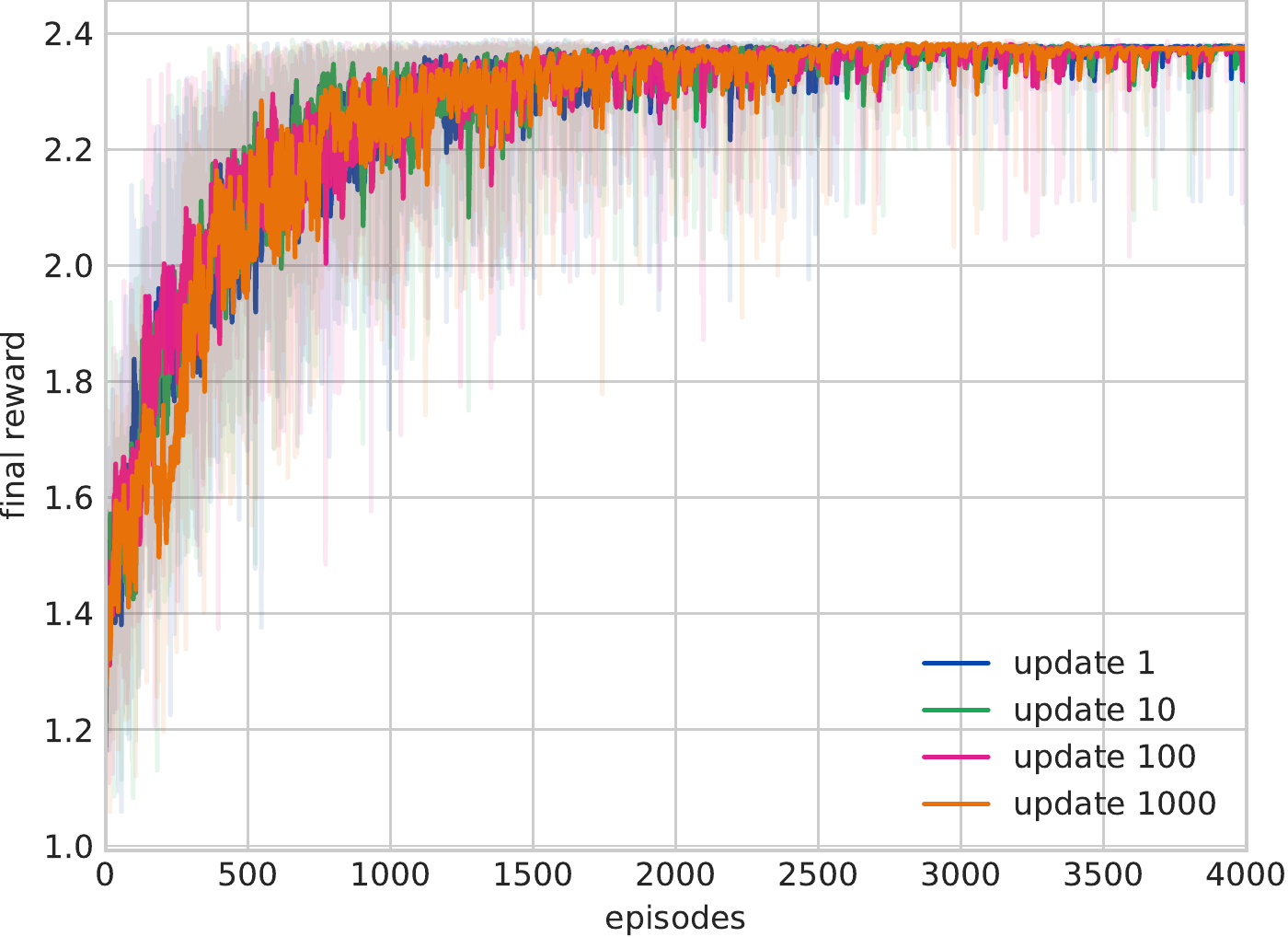}\hspace{-15pt}
			\label{subfig:updatefrequency}
		}
	\end{minipage}
	\begin{minipage}{0.3\textwidth}
		\subfigure[gamma]{
			\centering
			\includegraphics[width=1\textwidth]{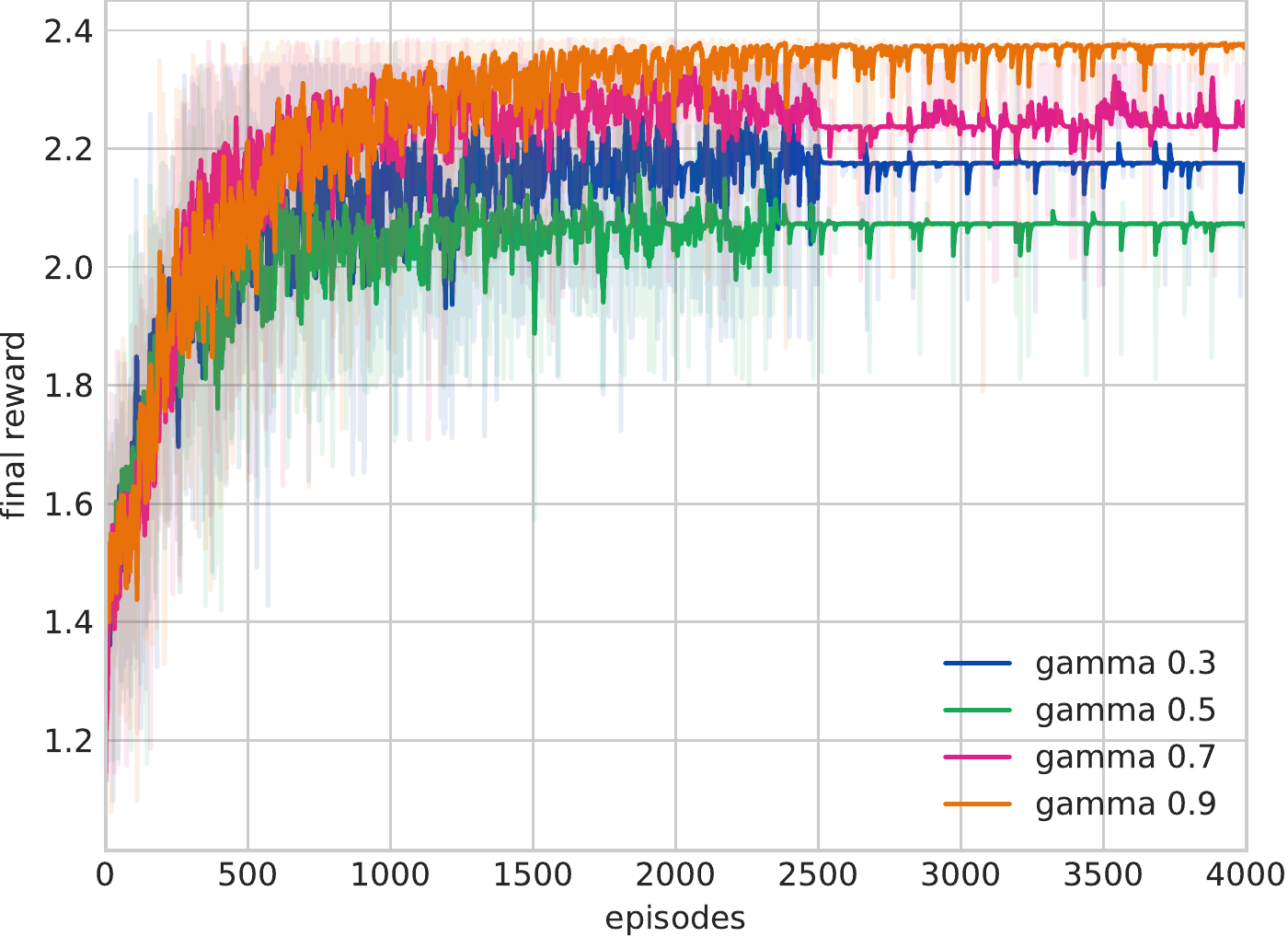}\hspace{-15pt}
			\label{subfig:gamma}
		}
	\end{minipage}	
	\begin{minipage}{0.3\textwidth}
		\subfigure[decay e-greedy]{
			\centering
			\includegraphics[width=1\textwidth]{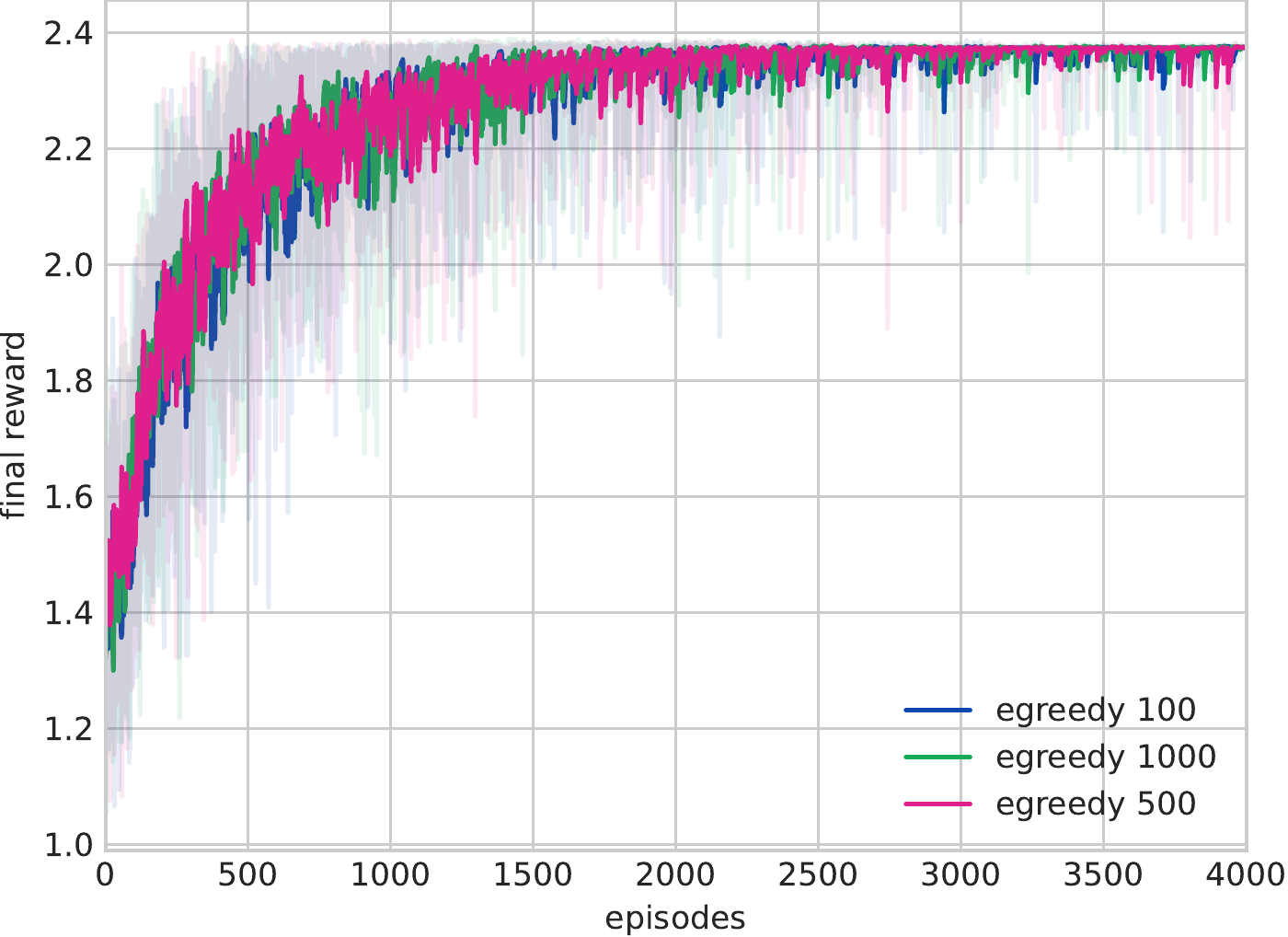}\hspace{-15pt}
			\label{subfig:e-greedy}
		}
	\end{minipage}	
	\caption{Different parameters setting.}
	\label{fig:param_drl}
\end{figure*}

The design of the reward function is the reason why the agent can obtain redundant links when constructing a multicast tree. It will select the action with the maximum Q value from all links according to the current state $S_t$, and selecting this link has little influence on the final reward value, which leads to the agent making decisions that are not conducive to the future.

The reason why it cannot converge to the optimal solution is that the agent will select the action that can complete the multicast tree before the state $S_{t+k}$ because the Q value of this action in the current state is larger than that of other edges, and it will not choose the longer but better link.

Exploring the influence of different hyperparameters on the convergence effect of the final reward value of an agent is discussed as follows.

The learning rate $\alpha$ determines the step size of the neural network parameter update, a larger setting leads to premature convergence or non-convergence of the model, and a setting too small will lead to slow convergence of the model and even lead to the model falling into local optimum. $\alpha$ is set to $10^{-4}$, $10^{-5}$, $10^{-6}$, and $10^{-7}$ to observe the convergence effect of the final reward value. Figure \ref{subfig:lr} shows that when $\alpha$ is $10^{-6}$ or $10^{-7}$, the agent falls into the local optimum; when $\alpha$ is $10^{-4}$ or $10^{-5}$, the agent searches in the direction of the global optimum.

The batch-size $k$ determines the size of each sample taken from the PER by an agent. Figure \ref{subfig:batchsize} shows that there is little influence on the overall curve of agent training. The final reward convergence is smoother with less fluctuation when $k$ is 8 or 16.

Figure \ref{subfig:nsteps} shows that the larger $n$ is, the easier it is to produce redundancy and even fall into a local optimum. When constructing a path in unicast, there is only one path to the destination node, and every decision on this path is an indispensable step after the future construction is completed. However, in the construction of a multicast tree, the link with the maximum Q value is selected from all links and added to the current tree at each decision. The reward value after n steps is calculated and amplified according to Formula \ref{eq:R_multistep}. This link may be redundant from the perspective of the whole tree in the future. Therefore, $n = 1$ is set.

The update frequency $n_{update}$ of the \textit{target network} is set to 1, 10, and 100. Figure \ref{subfig:updatefrequency} shows that when $n_{update}$ is set to the above value, there is little effect on the reward value.

The discount factor $\gamma$ is set to 0.3, 0.5, 0.7, and 0.9. Figure \ref{subfig:gamma} presents that the effect is best when $\gamma=0.9$. The agent cannot gain effective experience from the attempt, causing the agent to fall into a local optimum when $\gamma$ is small.

In Figure \ref{subfig:e-greedy}, the results show that there is little influence on curve convergence. Because the transitions generated by each decision are stored in the PER, the PER will preferentially select transitions with large TD errors when sampling a minibatch each time.

\subsection{Performance analysis}
\begin{figure*}[ht]
	\centering
	\begin{minipage}{0.24\textwidth}
		\subfigure[bandwidth]{
			\centering
			\includegraphics[width=1\textwidth]{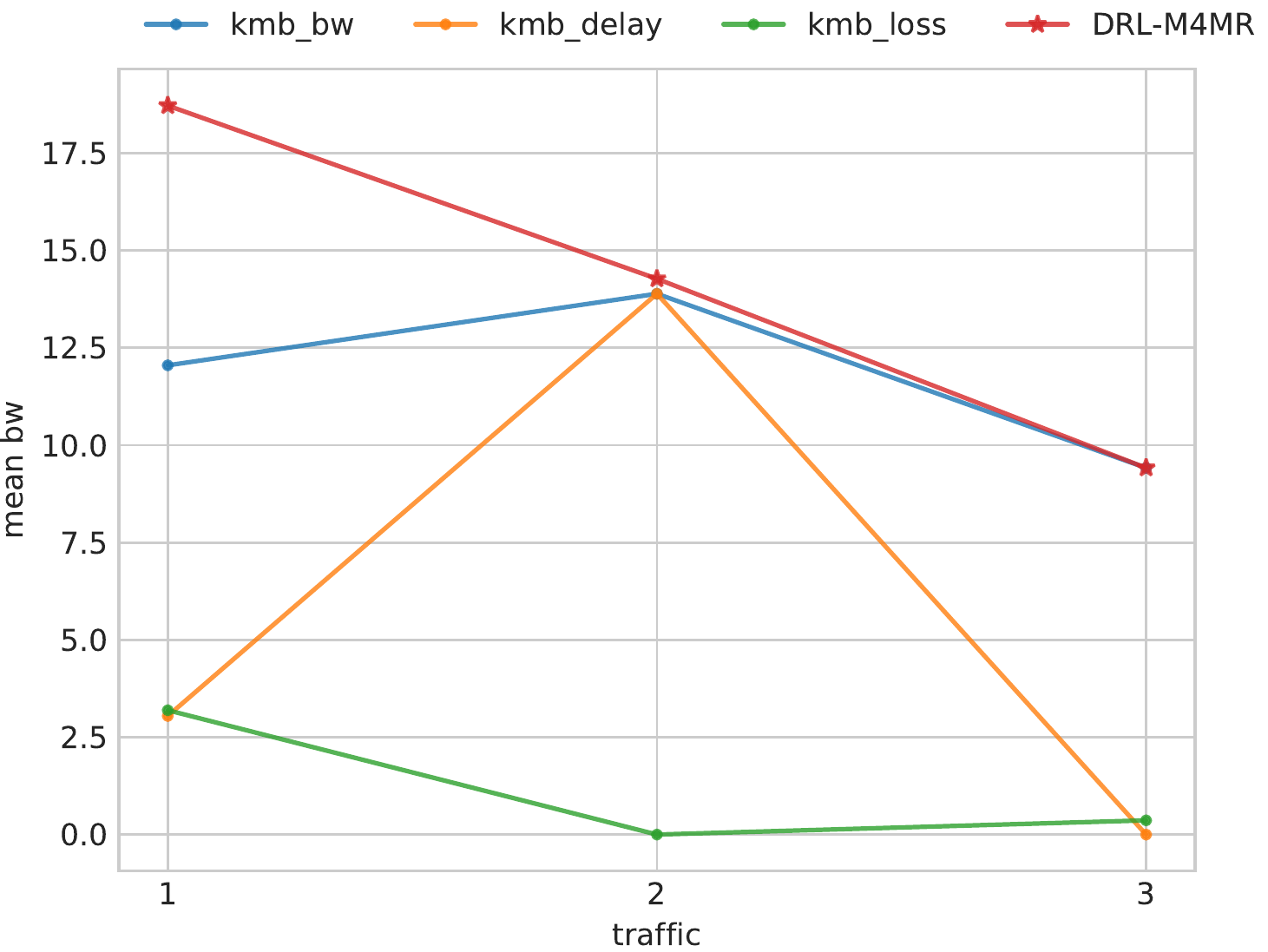}
			\label{subfig:compare_bw}
		}
	\end{minipage}
	\begin{minipage}{0.24\textwidth}
		\subfigure[delay]{
			\centering
			\includegraphics[width=1\textwidth]{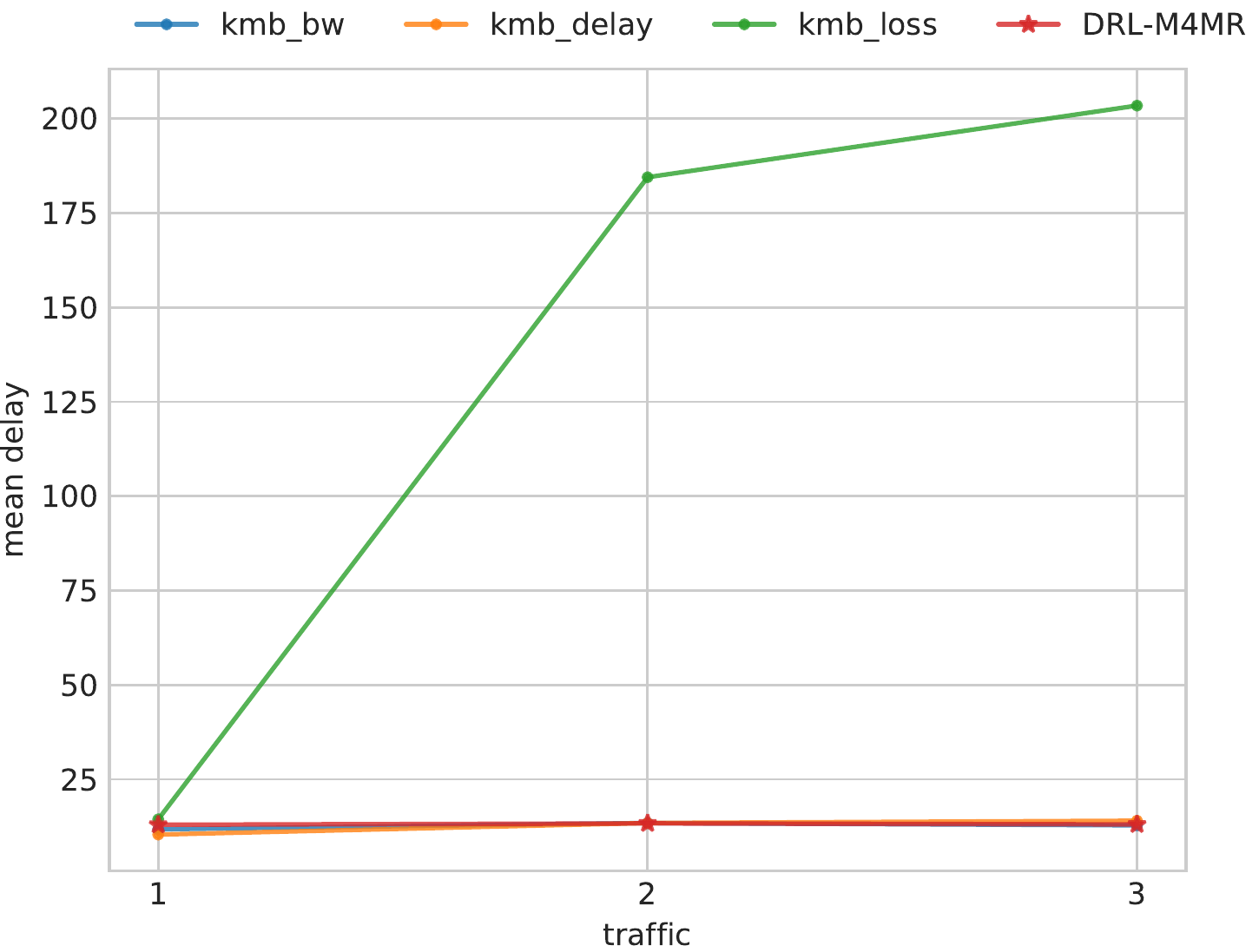}
			\label{subfig:compare_delay}
		}
	\end{minipage}
	\begin{minipage}{0.24\textwidth}
		\subfigure[loss]{
			\centering
			\includegraphics[width=1\textwidth]{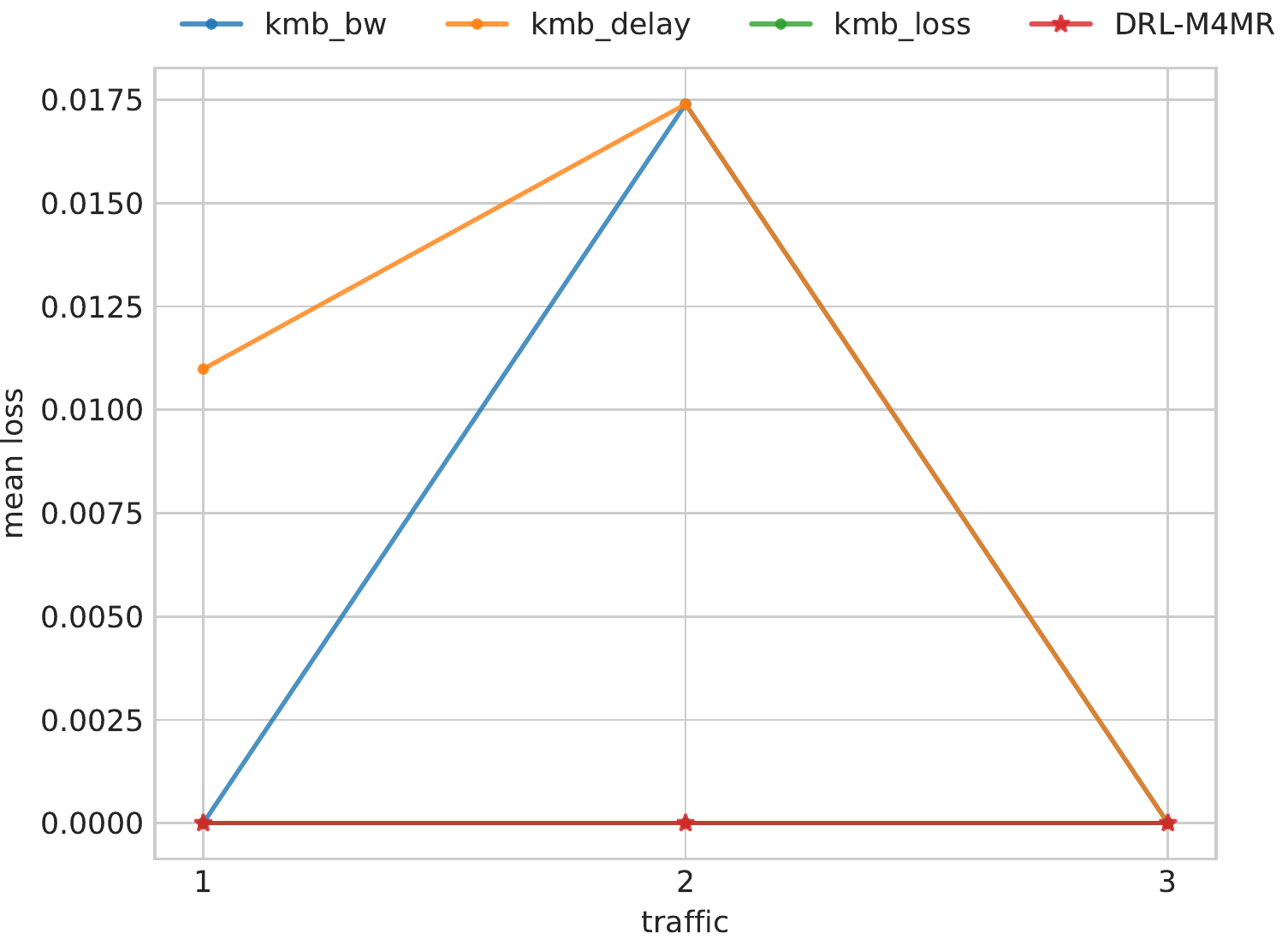}
			\label{subfig:compare_loss}
		}
	\end{minipage}
	\begin{minipage}{0.24\textwidth}
		\subfigure[length]{
			\centering
			\includegraphics[width=1\textwidth]{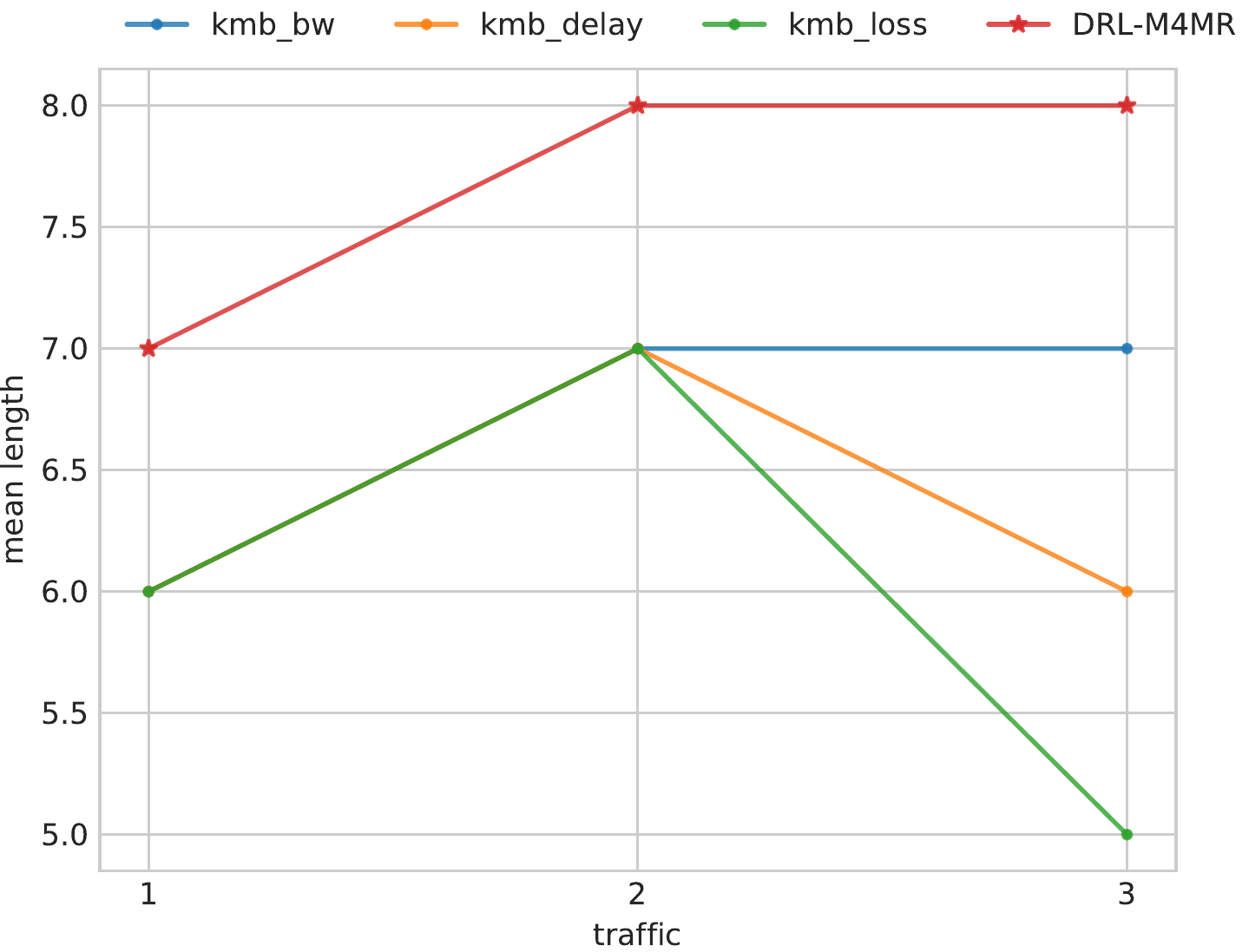}
			\label{subfig:compare_length}
		}
	\end{minipage}
	
	\caption{A comparison between DRL-M4MR and KMB algorithms.}
	\label{fig:comparison}
\end{figure*}

The DRL-M4MR is compared with the classical Steiner tree KMB algorithm. In the KMB algorithm, the residual bandwidth ($KMB_{bw}$), link delay ($KMB_{delay}$), and packet loss rate ($KMB_{loss}$) are used as weights for comparison. The three NLIs are used for DRL-M4MR agent training. After the training is completed according to Algorithm \ref{alg:DRL-M4MR}, the network parameter $\theta$ of the policy network is used for testing. The bandwidth, delay, packet loss rate, and link number of the multicast tree are compared, as shown in Figure \ref{fig:comparison}.

\begin{figure*}[h]
	\centering
	\begin{minipage}{1\textwidth}
		\subfigure[number=60]{
			\centering
			\includegraphics[width=0.24\textwidth]{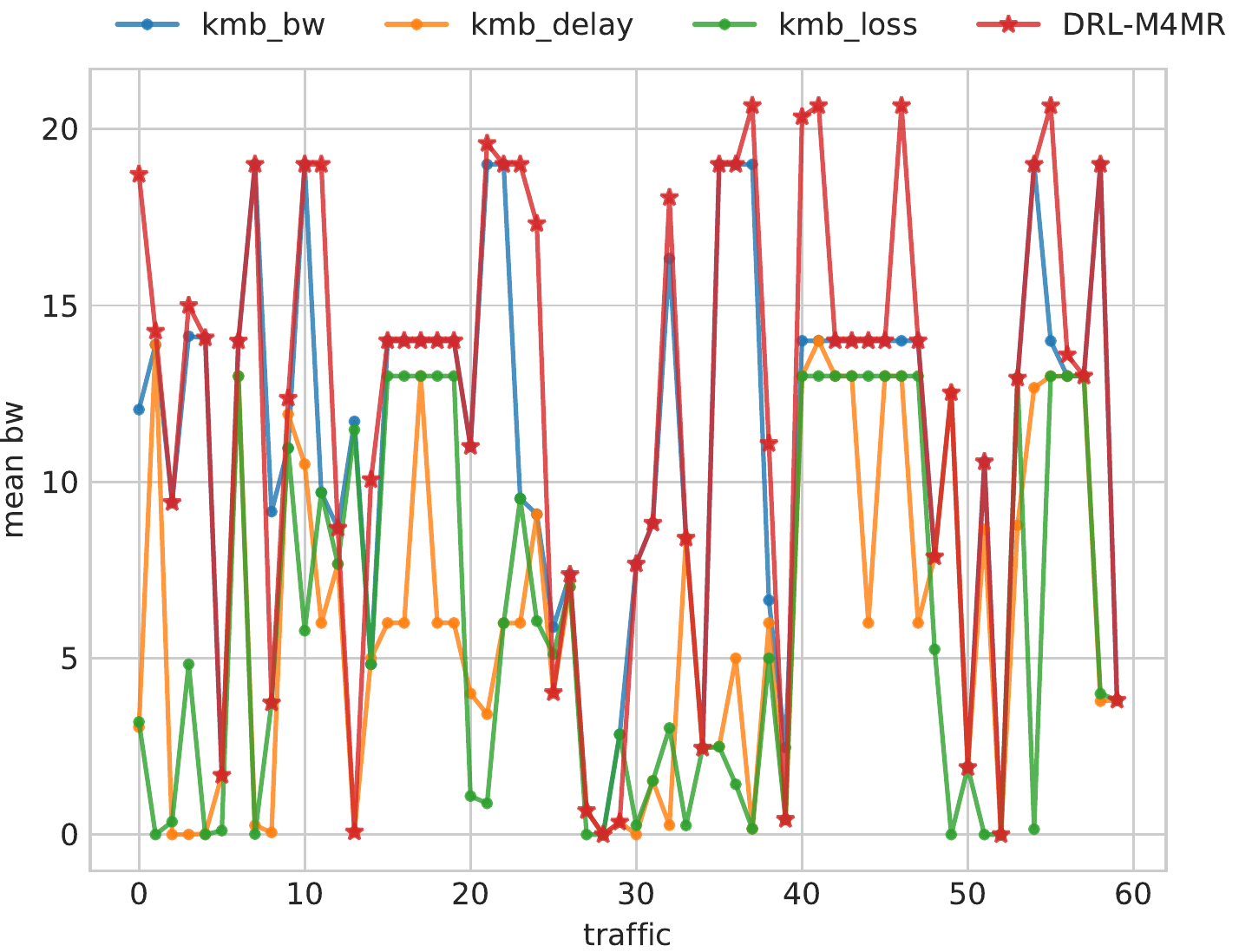}
			\includegraphics[width=0.24\textwidth]{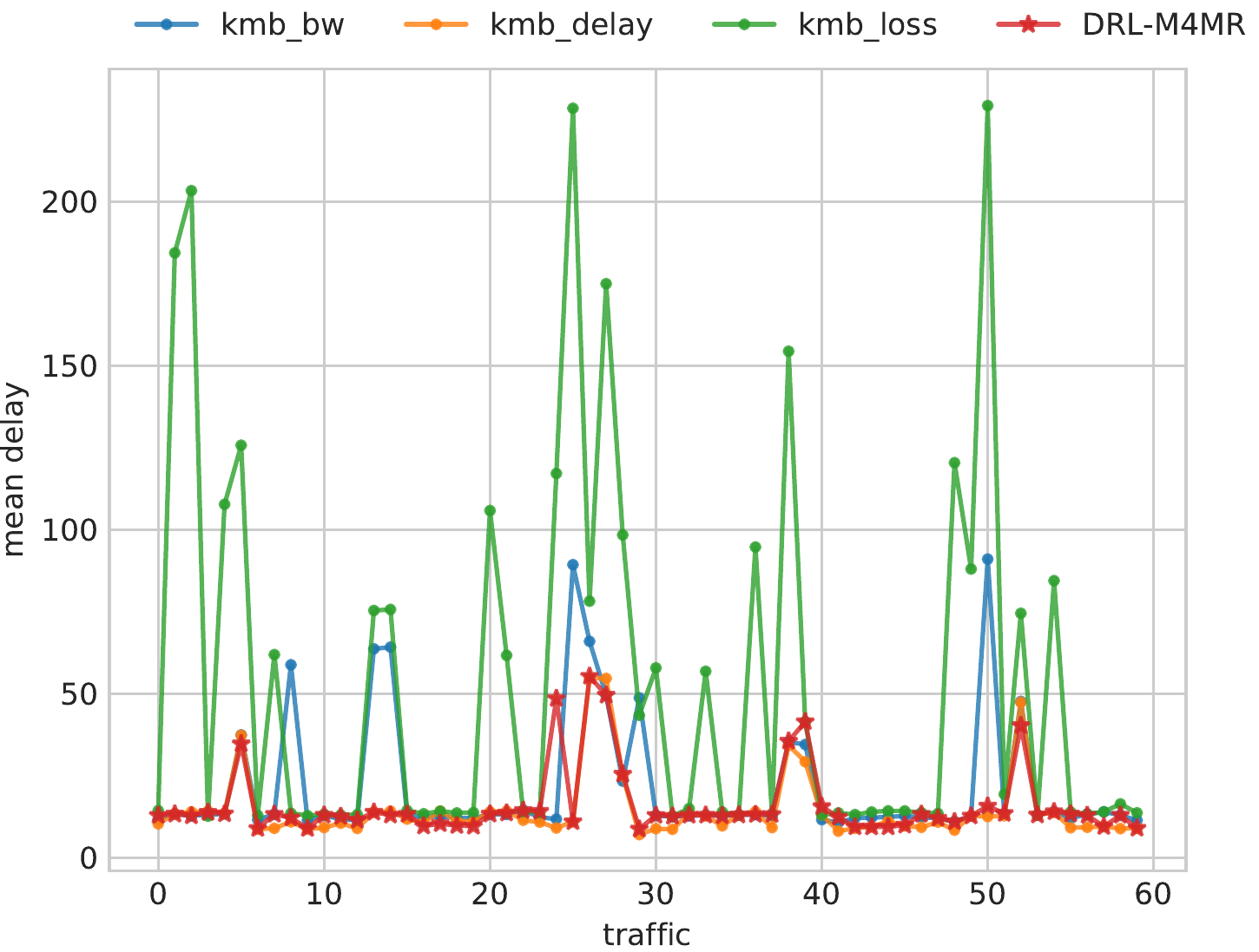}
			\includegraphics[width=0.24\textwidth]{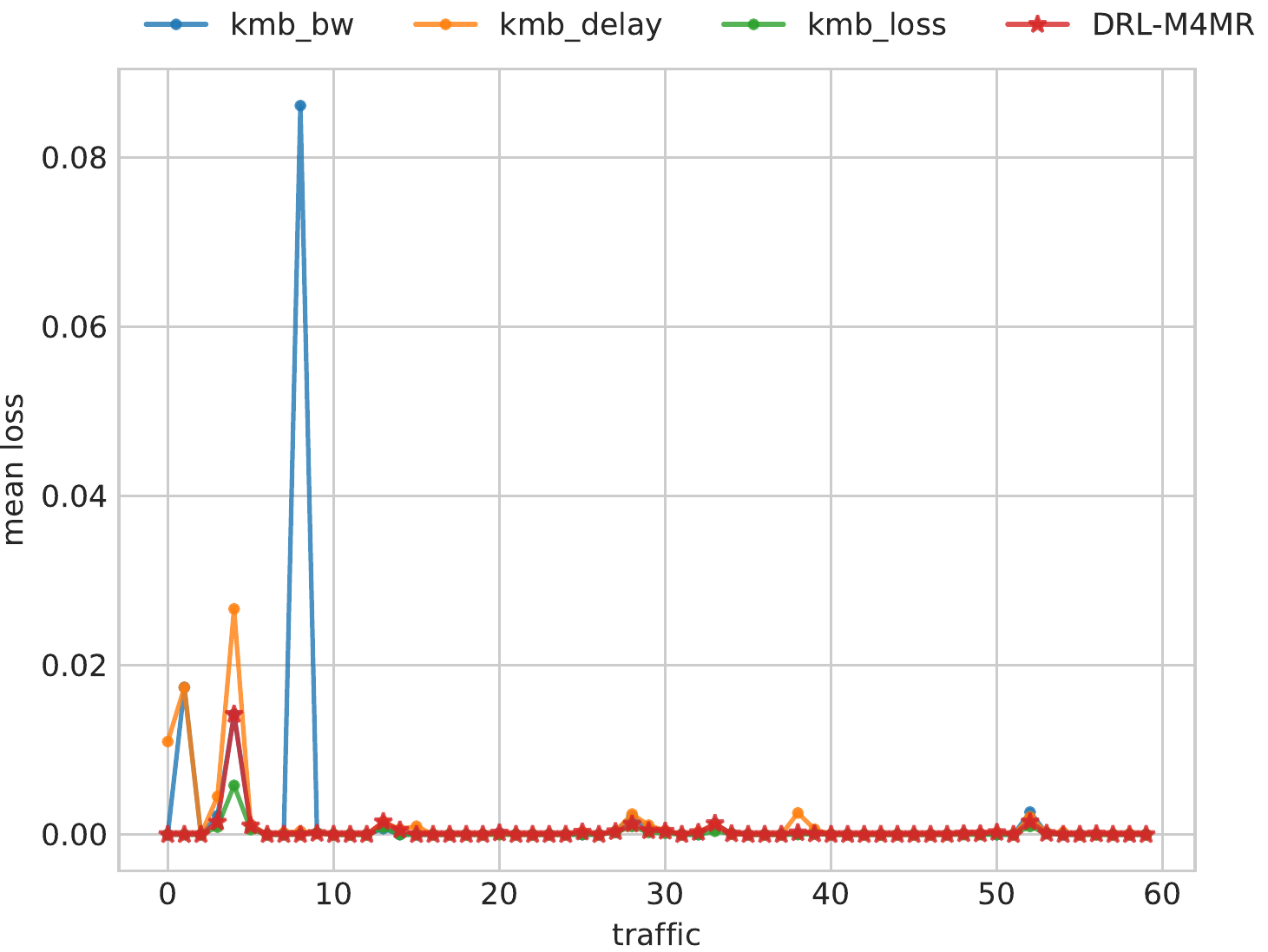}
			\includegraphics[width=0.24\textwidth]{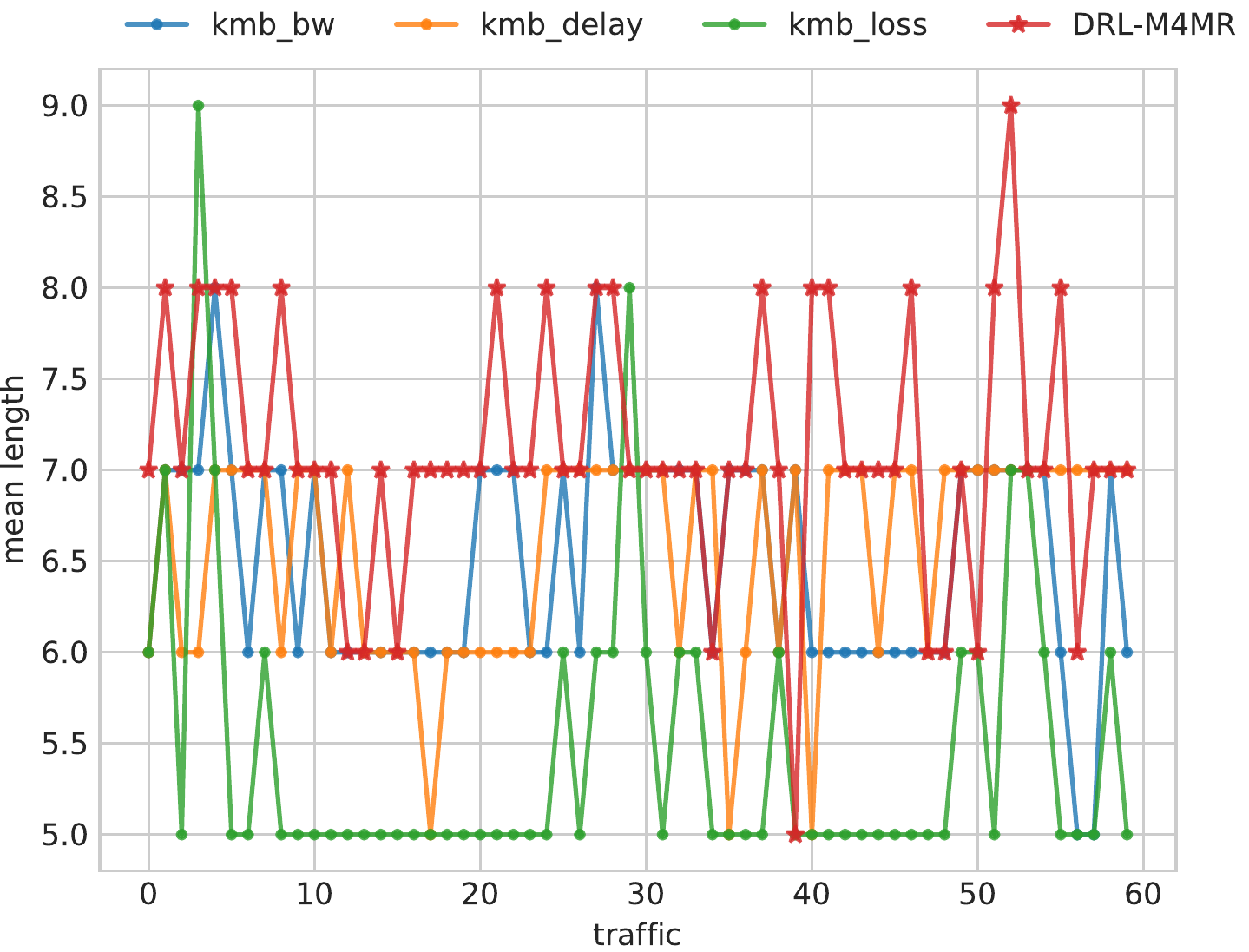}
			\label{subfig:number=60}
		}
	\end{minipage}
	\begin{minipage}{1\textwidth}
		\subfigure[number=120]{
			\centering
			\includegraphics[width=0.24\textwidth]{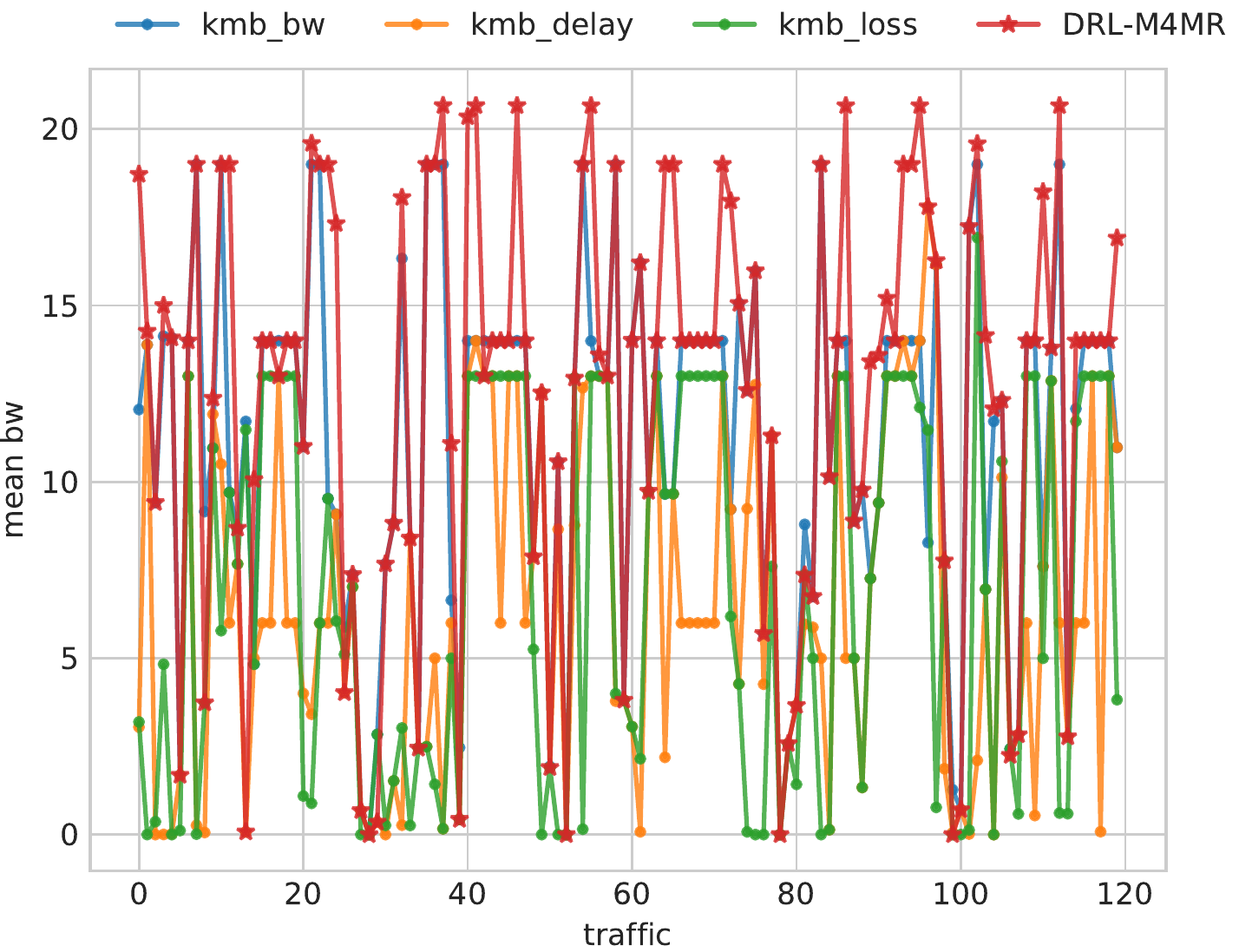}
			\includegraphics[width=0.24\textwidth]{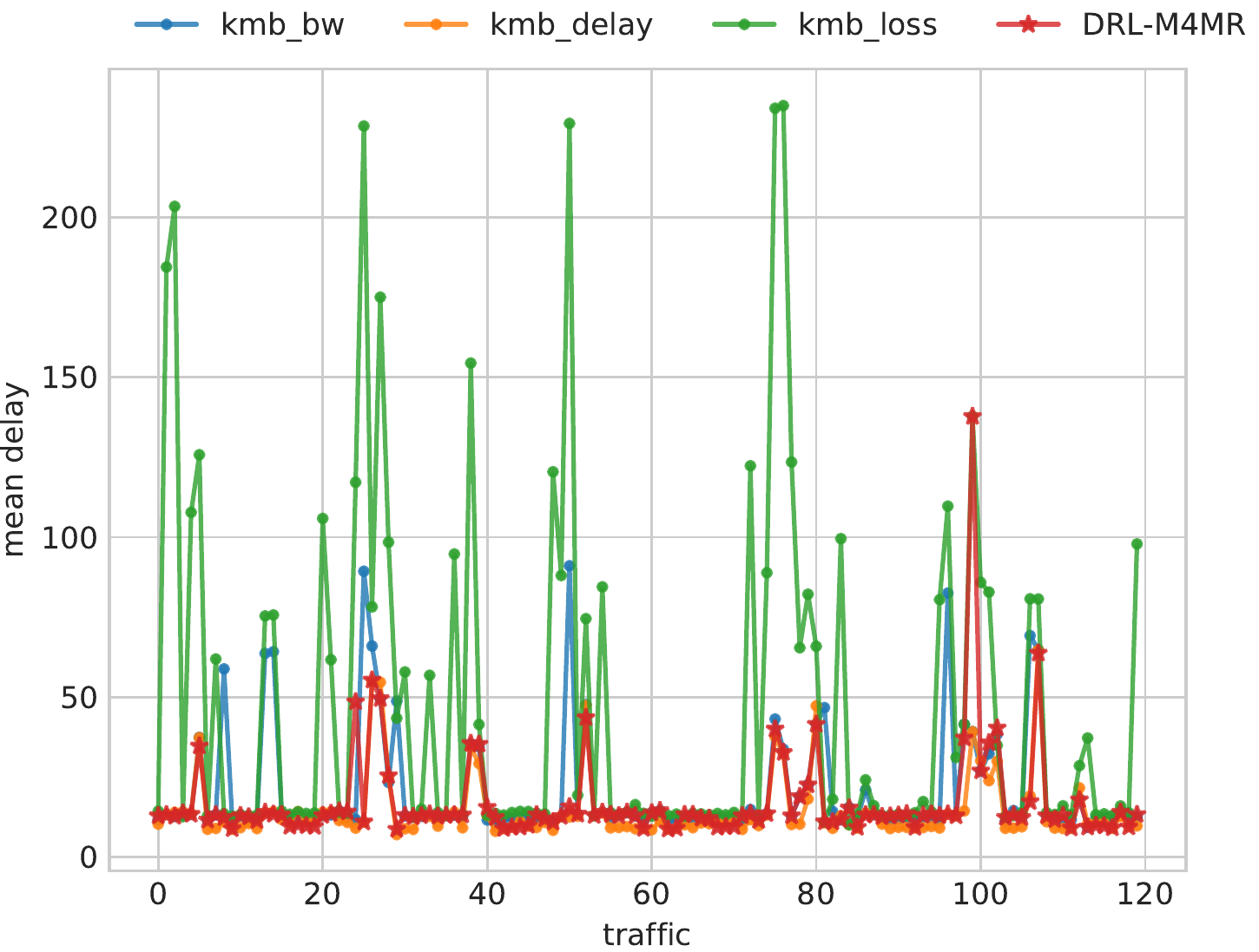}
			\includegraphics[width=0.24\textwidth]{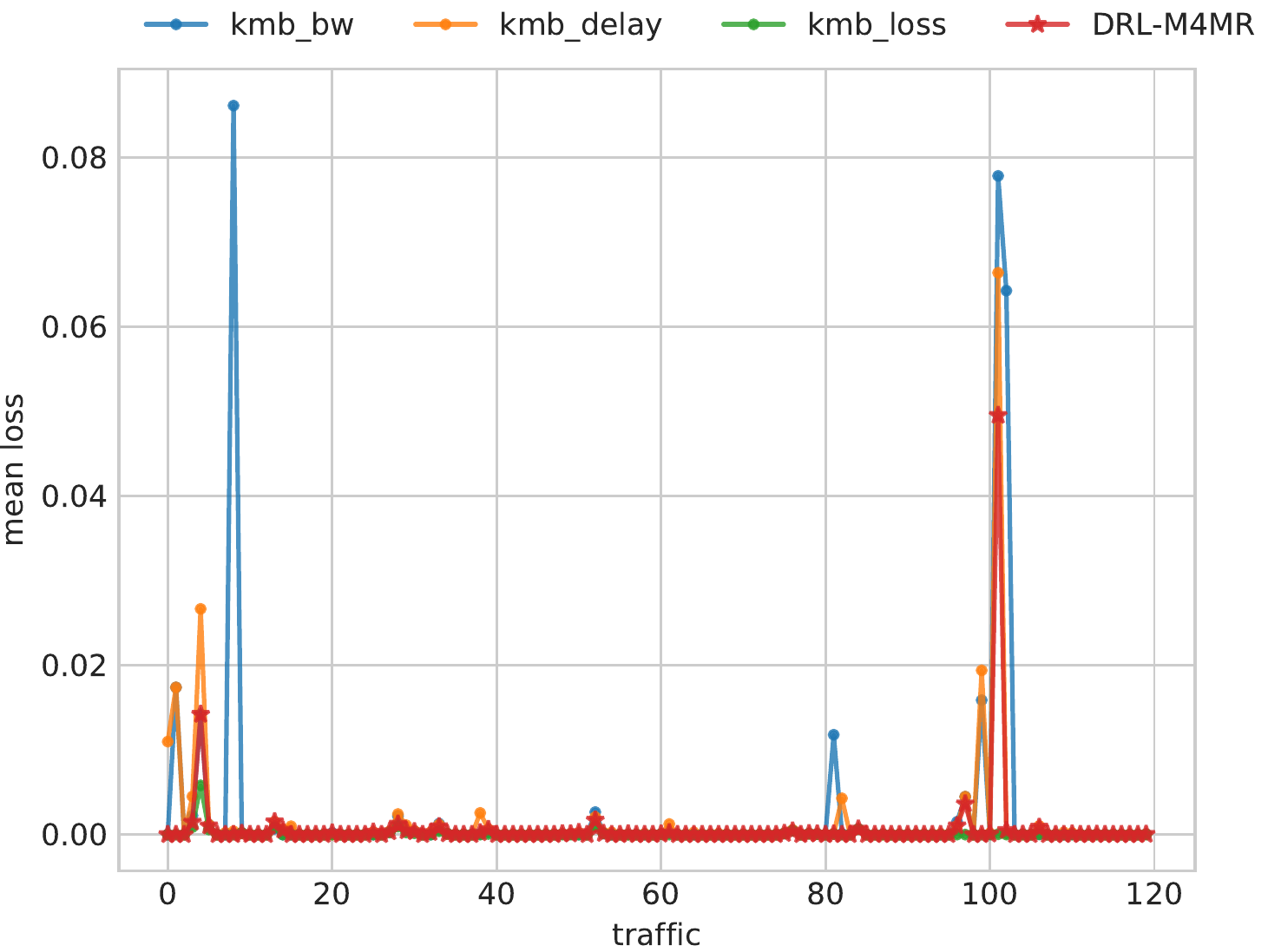}
			\includegraphics[width=0.24\textwidth]{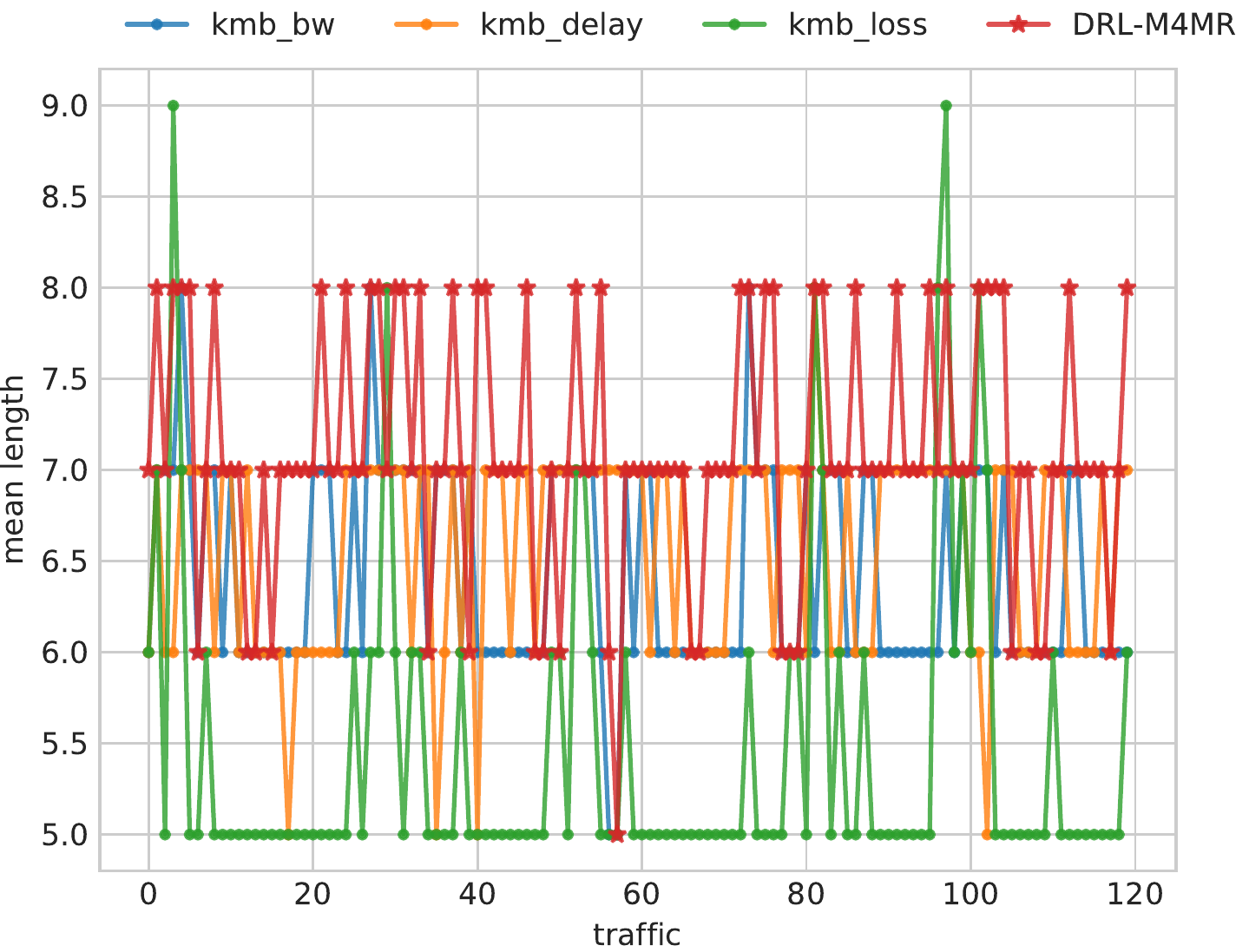}
			\label{subfig:number=120}	
		}
	\end{minipage}
	\caption{Different number of network link information}
	\label{fig:NLIs}
\end{figure*}

The metric in Figure \ref{subfig:compare_bw} is the average of the minimum remaining bandwidth of the path from the source node to each destination node in the multicast tree, that is, the average bottleneck bandwidth. The average bottleneck bandwidth produced by the DRL-M4MR algorithm is on average 19.37\% and at most 55.35\% higher than those produced by the $KMB_{bw}$ algorithm. The $KMB_{delay}$ and $KMB_{loss}$ algorithms show that the value of the average bottleneck bandwidth is 0 at worst, whereas the DRL-M4MR does not. The results show that the DRL-M4MR algorithm prefers to select the link with higher remaining bandwidth to construct a multicast tree.

Figure \ref{subfig:compare_delay} reveals that the average delay value of the multicast tree produced by the DRL-M4MR algorithm is on average 3.76\% and 5.8\% higher than $KMB_{bw}$ and $KMB_{delay}$, and 65.47\% lower than $KMB_{loss}$. At most, it is 10.24\% and 25.17\% higher than $KMB_{bw}$, $KMB_{delay}$, and 10.03\% lower than $KMB_{loss}$, respectively. Although the average delay produced by the DRL-M4MR algorithm is higher than that of $KMB_{delay}$, the values are very similar. The results show that the DRL-M4MR algorithm has a good delay performance.

In Figure \ref{subfig:compare_loss}, the metric is the average packet loss rate of multicast trees. The results show that the average packet loss rate produced by the DRL-M4MR algorithm on average is 33.33\% and 66.67\% lower than $KMB_{bw}$ and $KMB_{delay}$, respectively. The value is equal to those produced by $KMB_{loss}$, which indicates that the DRL-M4MR algorithm has a good packet loss rate performance.

Figure \ref{subfig:compare_length} presents that the link number of a multicast tree produced by the DRL-M4MR algorithm is on average 15.08\%, 21.43\% and 30.32\%, and at most 16.67\%, 33.33\% and 60.0\%, higher than the $KMB_{bw}$, $KMB_{delay}$ and $KMB_{loss}$ algorithms, respectively. The results indicate that DRL-M4MR selects a larger number of links to construct a multicast tree.

In the comparison of training of more NLIs, the training results under 60, and 120 NLIs are shown in Figure \ref{fig:NLIs}. The comparison results show that the agent of the DRL-M4MR can find the multicast tree with the maximum reward value according to the set reward function when constructing the multicast tree. 

Figure \ref{subfig:number=60} The bandwidth generated by the DRL-M4MR is on average 6.77\% higher than that of the $KMB_{bw}$ and much higher than that of the $KMB_{delay}$ and $KMB_{loss}$ algorithms. The delay is on average 5.67\%, 39.4\%, lower than $KMB_{bw}$ and $KMB_{loss}$, 18.31\% higher than $KMB_{delay}$. The packet loss rate is on average 1.97\%, 20.25\% lower than  $KMB_{bw}$ and $KMB_{delay}$, 17.71\% higher than $KMB_{loss}$. The length is on average 10.17\%, 9.32\%, 32.11\% higher than $KMB_{bw}$, $KMB_{delay}$ and $KMB_{loss}$. The results of Figure \ref{subfig:number=120} are similar to Figure \ref{subfig:number=60}.

The average bottleneck bandwidth is much better than that of the traditional KMB algorithm, the packet loss rate is also better, and the delay is not different from that of the traditional algorithm. However, for better transmission performance, multicast tree length is sacrificed, and the DRL-M4MR algorithm has the largest number of links in the multicast tree.

\begin{figure}[h]
	\centering
	\includegraphics[width=0.7\linewidth]{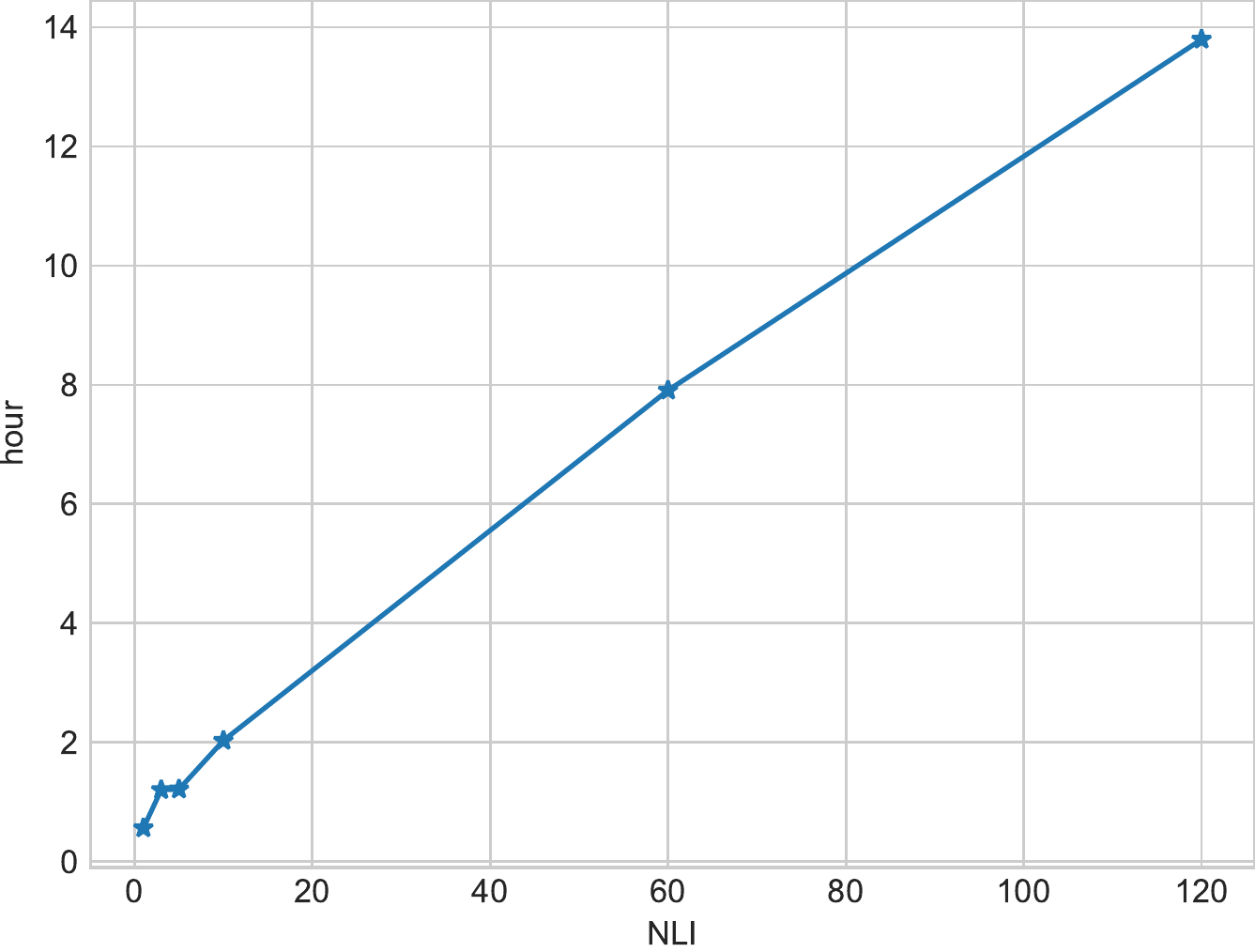}
	\caption{Training time cost with diferent number of NLIs.}
	\label{fig:time cost}
\end{figure}

Training episodes $M$ are set to 4000, and the cost of the training time with different numbers of NLIs are shown in Figure \ref{fig:time cost}. The results show that the larger the number of NLIs, the longer the training time, in a linear increasing relationship.

\section{Conclusion}\label{sec:Conclusion}
In this paper, a multicast routing method DRL-M4MR based on DRL in SDN is proposed. In a dynamic SDN environment, DRL-M4MR utilizes the double dueling DQN reinforcement learning method. After training, the agent can construct a multicast tree with optimal metrics according to the measured network link information. The experimental results show that in most cases, DRL-M4MR can construct multicast trees with a higher average bottleneck bandwidth, a lower average delay, and a lower average packet loss rate but more links. When the traffic in the data plane changes, the agent can intelligently adjust the multicast tree to make the optimal metrics of the multicast tree in the current network link information, and the performance of data transmission is guaranteed.

For future work, first, we intend to improve the reward mechanism. The design of the step reward and final reward cause redundant branches; if only the final reward is used to measure the quality of the multicast tree, it will cause the agent to fall into the local optimum, so hierarchical reinforcement learning will be considered in the future. Second, distributed reinforcement learning is considered to improve training efficiency and reduce training time.

\section*{Acknowledgement}
This work obtained the subsidization of National Natural Science Foundation of China (No.62161006), Innovation Project of Guangxi Graduate Education (No. YCSW2022271), Guangxi Natural Science Foundation of China (No. 2018GXNSFAA050028), and Guangxi Key Laboratory of Wireless Wide band Communication and Signal Processing (No. GXKL06220110).



  \bibliographystyle{elsarticle-num} 
  \bibliography{elsarticle}




%
\end{document}